\documentclass[12pt,english]{article}

\usepackage[T1]{fontenc}
\usepackage[utf8]{inputenc}
\usepackage{geometry}
\usepackage{amsmath}
\usepackage{amsfonts}
\usepackage{amssymb}
\usepackage{graphicx}
\usepackage{setspace}
\usepackage{slashed,ulem}

\usepackage{xcolor} % added here

\makeatletter

\providecommand{\tabularnewline}{\\}

\newcommand{\lyxaddress}[1]{
	\par {\raggedright #1
	\vspace{1.4em}
	\noindent\par}
}

\usepackage[winfonts,UTF8]{ctex}

\usepackage{slashed}
\usepackage{hyperref}
\usepackage{graphicx}
\usepackage[numbers,sort&compress]{natbib}
\date{}

\hypersetup{colorlinks=true,citecolor=blue,linkcolor=red}

\makeatother

\usepackage{babel}
\begin{document}
\title{Dynamical charged black hole spontaneous scalarisation
in Anti-de Sitter spacetimes}
\author{Cheng-Yong Zhang$^{1}$, Peng Liu$^{1}$, Yunqi Liu$^{2}$, Chao
Niu$^{1}$,Bin Wang$^{2,3}$ \thanks{zhangcy@email.jnu.edu.cn, phylp@jnu.edu.cn, yunqiliu@yzu.edu.cn, niuchaophy@gmail.com,
wang\_b@sjtu.edu.cn}}
\maketitle

\lyxaddress{\begin{center}
\textit{1. Department of Physics and Siyuan Laboratory, Jinan University,
Guangzhou 510632, China}\\
\textit{2. Center for Gravitation and Cosmology, College of Physical
Science and Technology, Yangzhou University, Yangzhou 225009, China}\\
\textit{3. School of Aeronautics and Astronautics, Shanghai Jiao Tong
University, Shanghai 200240, China}
\par\end{center}}
\begin{abstract}
We study the fully nonlinear dynamics of black hole spontaneous scalarizations  in Einstein Maxwell scalar theory with coupling function $f(\phi)=e^{-b\phi^{2}}$, which can transform usual Reissner-Nordström Anti-de Sitter (RN-AdS) black holes into hairy black holes.  Fixing the Arnowitt-Deser-Misner mass of the system, the initial scalar perturbation will destroy the original RN-AdS black hole and turn it into a hairy black hole provided that the constant  $-b$ in the coupling function and the charge of the original black hole are sufficiently large, while the cosmological constant is small enough. In the scalarization process, we observe that the black hole irreducible mass initially increases exponentially, then it approaches to and finally saturates at a finite value. Choosing stronger coupling and larger black hole charge, we find that the black hole mass exponentially grows earlier and it takes longer time for a hairy black hole to be developed and stabilized.  We further examine phase structure properties in the scalarization process and confirm the observations in the non-linear dynamical study.
\end{abstract}

\section{Introduction }

Though general relativity is mathematically beautiful and experimentally tested correct up to now, it may be not a perfect theory of gravity. It requires unknown physics such as dark energy and dark matter to describe the current universe and contains the non-renormalizable curvature singularity.  To overcome these drawbacks, generalised theories of gravity are called for.
In general relativity, there is a black hole  ``no-hair'' theorem claiming that a black hole can only be determined by its mass, charge and angular-momentum \cite{Israel1967,Carter1971,Ruffini1971} and one cannot learn more information except for these.  However, it was observed that for black hole solutions with a Yang-Mills  \cite{HHVolkov1989,HHBizon1990,HHGreene1993,HHMaeda1994} or Skyrme field \cite{HHLuckock1986,HHDroz1991} or a conformally-coupled scalar field \cite{HHBekenstein1975}, the no-hair theorem was evaded.
A novel no-hair theorem was proposed for non-minimally coupled scalar field \cite{HHBekenstein1995} and was extended to general scalar-tensor theories \cite{Sotiriou:2011dz,Hui:2012qt}.
But it was again found violated in the case of dilatonic and colored black holes in the context of the Einstein-dilaton-Gauss-Bonnet theory \cite{Torii1997,Kanti1997}. The rotating \cite{EdGBP1,EdGBP2,EdGBP3} or higher dimensional \cite{EdGBP4,EdGBP5,EdGBP6,EdGBP7,EdGBP8,EdGBP9} or shift-symmetric Galileon \cite{EdGBP10,EdGBP11,EdGBP12} hairy black hole solutions were later constructed.

There appears some dynamical mechanisms leading to hairy black hole solutions, one of which is the spontaneous scalarization which has attracted a lot of attentions recently.
This mechanism is typically introduced by a non-minimal coupling between scalar field and some source term which can trigger a repulsive gravitational effect, via an effective tachyonic mass for the scalar. The tachyonic instability induces an exponential growth of the scalar field, and leads to hairy black hole solutions.
The spontaneous scalarisation was first proposed in the study of the neutron
star in scalar tensor theory by  coupling scalar fields to the Ricci curvature \cite{Damour1993}.
It was also found taking place in scalar tensor theory if black holes are surrounded by  sufficient
amount of matter \cite{Cardoso1305,Cardoso1308,Zhang:2014kna}. Recent works found
that the spontaneous scalarisation can be triggered by the geometric invariant source  such as the Gauss-Bonnet invariant in extended Scalar-Tensor-Gauss-Bonnet (eSTGB) theory \cite{Doneva1711,Silva1711,Antoniou1711,Cunha1904,Dima:2020yac,Herdeiro2009,Berti2009}, the Ricci scalar for non-conformally invariant black holes \cite{Herdeiro:2019yjy}, or the Chern-Simons invariant \cite{Brihaye:2018bgc}. The spontaneous scalarisation can also be triggered by the matter invariant source such as  in Einstein-Maxwell-scalar
(EMS) theory \cite{Herdeiro:2018wub} and Einstein-Maxwell-vector
model  \cite{Oliveira:2020dru}. More recent works on spontaneous scalarisations of compact objects in gravity can be found in references  \cite{Brihaye:2019puo,Astefanesei:2019pfq,Peng:2019qrl,Astefanesei:2020qxk,Blazquez-Salcedo:2018jnn,Macedo:2019sem,Lin:2020asf,Guo:2020sdu,Bakopoulos:2020dfg,Fernandes:2020gay,Collodel:2019kkx,Brihaye:2019dck,Herdeiro:2019iwl,Doneva:2018rou,Blazquez-Salcedo:2020nhs,Guo:2021zed}.

The stability of the hairy black hole solution depends on the coupling
function and ranges of parameters in the system. For examples,
scalarised black holes are thermodynamically favored and  fundamental
branches of scalarised black holes are stable
against perturbations in eSTGB theory with exponential-type
coupling functions \cite{Doneva1711}. While scalarised solutions
are not thermodynamically favored and fundamental branches evolve
unstably against perturbations in eSTGB theory with power-law
coupling function \cite{Silva1711}.
Most available works focused on linear stability discussions
of hairy black holes \cite{Myung:2018vug,Myung:2018jvi,Myung:2019oua,Zou:2019bpt,Myung:2020ctt,Silva:2018qhn,Hod:2019pmb,Konoplya:2019fpy,Blazquez-Salcedo:2020rhf,Zhang:2020pko}. But the endpoint of the stability can only be determined once fully non-linear numerical studies of perturbation evolution are carried out.  In eSTGB theory, the numerical studies on the gravitational collapse and the evolution of the black hole reveal that stable hairy black holes can  be formed as the final state  \cite{Ripley:2019irj,Ripley:2019aqj,Ripley:2020vpk,Doneva:2021dqn}. However, the equations of motion in eSTGB theory are well-posed only when the Gauss-Bonnet coupling parameter is sufficiently small, otherwise, the dynamical studies can encounter technical difficulty.
% and makes the dynamical studies technical difficult in general.
% On the other hand,
In EMS theory, we find that different from eSTGB theory, the dynamic equation is well-posed for large couplings. It is expected that EMS theory allows fully non-linear dynamical evolution
and thus can reflect the detailed process of the black hole spontaneous scalarisation. We expect to disclose the growth of the scalar hair and the examine irreducible mass of the black hole to get deeper insights into  physics on the scalarization process.

Some non-linear works have been done in the
asymptotic flat spacetime \cite{Herdeiro:2018wub,Fernandes:2019rez,Fernandes:2019kmh}.
The cosmological constant was argued playing the role in
the black hole scalarisation. The regular hairy
black hole solutions were found in all asymptotic flat, de Sitter (dS) and AdS spacetimes in EMS
models \cite{Brihaye:2019gla}. In eSTGB model, hairy black holes were found existing in asymptotic flat and AdS spacetimes
\cite{Bakopoulos:2018nui}, but not in dS spacetime since it was argued that the positive cosmological constant can quench the tachyonic instability.
Most of discussions on the cosmological constant effect in scalarizations are limited in linear perturbation levels. In this paper, we focus on the fully non-linear
dynamical evolution of black holes in asymptotic AdS spacetimes in EMS theory. The dynamics in AdS spacetime is significantly
different from those in asymptotically flat spacetime since the scalar
wave can reach the spacial boundary at finite coordinate and
be bounced back. As we will see later, the studies in AdS spacetime can not be generalized straightforwardly to the case of asymptotically flat spacetime  since their boundary conditions are completely different.
The spontaneous scalarization in the AdS spacetime is in the spirit similar to the spontaneous condensation in holographic theories \cite{Hartnoll:2008kx,Cai:2015cya}. Besides disclosing the dynamical phenomenon in the bulk nonlinear perturbations, we can borrow the idea of studying phase transitions in holographic superconductors to relate the critical point of phase transitions to an eigenvalue problem to examine  detailed critical lines separating scalarized hairy black holes from original RN-AdS black holes. We examine phase structures of the system and find that physics obtained matches exactly to our dynamical calculation results. In the phase structure study we confirm that the scalarization occurs more easily with larger values of $Q$ and smaller values of $-\Lambda$.

This paper is organized as follows. In section 2, we introduce
equations of motions and boundary behaviors of the variables in EMS
theory. In section 3, we show the numerical results and illustrate the effects of
the coupling function parameter, the charge and the cosmological constant on the dynamical scalarisation. Further we disclose properties in phase structures and confirm the findings in non-linear dynamical systems.  Section 4 gives a summary and discussion on the obtained results.

\section{Model}

The action for the Einstein-Maxwell-scalar theory with a negative cosmological constant $\Lambda$ reads
\begin{equation}
S=\frac{1}{16\pi}\int d^{4}x\sqrt{-g}\left[R-2\Lambda-2\nabla_{\mu}\phi\nabla^{\mu}\phi-f(\phi)F_{\mu\nu}F^{\mu\nu}\right].
\end{equation}
 Here $R$ is the Ricci scalar, $\phi$ is a real scalar, $A_{\mu}$ is the Maxwell field and its strength $F_{\mu\nu}=\partial_{\mu}A_{\nu}-\partial_{\nu}A_{\mu}$.
 In this paper, we consider the coupling function as $f(\phi)=e^{-b\phi^{2}}$ with $b$ a dimensionless coupling constrant.
 This model admits both Reissner-Nordström (RN)-AdS solutions and hairy black hole solutions \cite{Guo:2021zed}.
 Varying the action with respect to $g^{\mu\nu}$, $A^{\mu}$ and $\phi$, we obtain the equations of motion for the fields in our model.
 The equations of motion for gravity are
\begin{equation}\label{gra}
R_{\mu\nu}-\frac{1}{2}Rg_{\mu\nu}+\Lambda g_{\mu\nu}=2\left(T_{\mu\nu}^{\phi}+f(\phi)T_{\mu\nu}^{A}\right),
\end{equation}
with the energy momentum tensor
\begin{eqnarray}
T_{\mu\nu}^{\phi}&=&\partial_{\mu}\phi\partial_{\nu}\phi-\frac{1}{2}g_{\mu\nu}\nabla_{\rho}\phi\nabla^{\rho}\phi, \\
T_{\mu\nu}^{A}&=&F_{\mu\rho}F_{\nu}^{\ \rho}-\frac{1}{4}g_{\mu\nu}F_{\rho\sigma}F^{\rho\sigma}.
\end{eqnarray}
The equation of motion for the real scalar field $\phi$ is
\begin{equation}\label{scalarequation}
\nabla_{\mu}\nabla^{\mu}\phi=\frac{1}{4}\frac{df(\phi)}{d\phi}F_{\mu\nu}F^{\mu\nu}.
\end{equation}
In eq.\ref{scalarequation} the term on the right hand side is related to the coupling function, this term could be considered as an effective mass term which is essential for the black hole spontaneous scalarisation \cite{Herdeiro:2018wub,Brihaye:2019gla}.
The equations for the gauge field $A_{\mu}$ are given by
\begin{equation}\label{electric-mag}
\nabla_{\mu}\left(f(\phi)F^{\mu\nu}\right)=0.
\end{equation}

\subsection{Equations of motion and boundary conditions}
We  study the dynamic process of the spontaneous scalarisation of spherically symmetric black holes in EMS theory.
The ansatz of the spacetime in the spherical symmetry could takes as the ingoing Eddington-Finkelstein coordinate which is regular on the black hole horizon,
\begin{equation}
ds^{2}=-\alpha(t,r)dt^{2}+2dtdr+\zeta(t,r)^{2}(d\theta^{2}+\sin^{2}\theta d\phi^{2}).
\end{equation}
We choose the gauge to set the ansatz of gauge field as $A_{\mu}dx^{\mu}=A(t,r)dt$, and the scalar field behaves as $\phi=\phi(t,r)$.

From eq.(\ref{electric-mag}) and the ansatz, the equations for the gauge field reduces to
\begin{equation}\label{em}
\partial_{r}\left(\zeta^{2}f(\phi)E\right)=0,\ \ \ \partial_{t}\left(\zeta^{2}f(\phi)E\right)=0,
\end{equation}
with $E\equiv\partial_{r}A$.
The solution to eq.(\ref{em}) can be written as
\begin{equation}
E=\frac{Q}{\zeta^{2}f(\phi)},
\end{equation}
where $Q$ is a constant interpreted as the electric charge.

Plugging the ansatz into eq.(\ref{gra}) and using the auxiliary variables eq. (\ref{aux-var}), the equations of motion for metric are written as
\begin{eqnarray}\label{eqall}
\partial_{t}S&=&\frac{1}{2}S\partial_{r}\alpha+\frac{\alpha}{2}\left(\frac{2S\partial_{r}\zeta-1}{2\zeta}+\frac{1}{2}\zeta\Lambda+\frac{Q^{2}}{2\zeta^{3}f(\phi)}\right)-\zeta P^{2},\label{eq:St}\\
\partial_{r}^{2}\alpha&=&-4P\partial_{r}\phi+\frac{4S\partial_{r}\zeta-2}{\zeta^{2}}+\frac{4Q^{2}}{\zeta^{4}f(\phi)},\label{eq:alphar}\\
\partial_{r}S&=&\frac{1-2S\partial_{r}\zeta}{2\zeta}-\frac{\zeta\Lambda}{2}-\frac{Q^{2}}{2\zeta^{3}f(\phi)},\label{eq:Sr}\\
\partial_{r}^{2}\zeta&=&-\zeta(\partial_{r}\phi){}^{2},\label{eq:zetar}
\end{eqnarray}
the scalar equation becomes
\begin{equation}
\partial_{r}P=-\frac{P\partial_{r}\zeta+S\partial_{r}\phi}{\zeta}-\frac{Q^{2}}{4\zeta^{4}f(\phi)^{2}}\frac{df(\phi)}{d\phi}.\label{eq:Pr}
\end{equation}
To implement the numerical method, the auxiliary variables \cite{Bosch:2016vcp} are introduced as,
\begin{eqnarray}\label{aux-var}
S & =\partial_{t}\zeta+\frac{1}{2}\alpha\partial_{r}\zeta,\label{eq:S}\\
P & =\partial_{t}\phi+\frac{1}{2}\alpha\partial_{r}\phi.\label{eq:Pt}
\end{eqnarray}

\subsection{Boundary conditions and numerical method}\label{Boundary}
We expect the information on the solutions with non-trivial scalar field, in this section we firstly specify the boundary conditions.
Expanding the variables $\phi$, $\alpha$  and $\zeta$ at spatial infinity, substituting the series into eqs.(\ref{eq:St}-\ref{eq:Pr}) and solving the equations order by order,
 we get the power series expansions of the variables near the spatial infinity:
\begin{eqnarray}\label{expa}
\phi(t,r)  =& \frac{\phi_{3}(t)}{r^{3}}+\frac{3}{8\Lambda r^{4}}\left(\frac{Q^{2}f'(0)}{f(0)^{2}}-8\phi'_{3}(t)\right)+O(r^{-5}),\label{phiexp}\\
\alpha(t,r)  =& -\frac{\Lambda}{3}r^{2}+1-\frac{2M}{r}+\frac{Q^{2}}{f(0)r^{2}}+\frac{\Lambda}{5r^{4}}\phi_{3}^{2}(t)+O(r^{-5}),\\
\zeta(t,r)   =& r-\frac{3\phi_{3}^{2}(t)}{10r^{5}}+\frac{3\phi_{3}(t)}{14\Lambda r^{6}}\left(\frac{Q^{2}f'(0)}{f(0)^{2}}-8\phi'_{3}(t)\right)+O(r^{-7}),\\
S(t,r)  = & -\frac{\Lambda}{6}r^{2}+\frac{1}{2}-\frac{M}{r}+\frac{Q^{2}}{2f(0)r^{2}}-\frac{3\Lambda}{20r^{4}}\phi_{3}^{2}(t)+O(r^{-5}),\\
P(t,r)  =& \frac{\Lambda\phi_{3}(t)}{2r^{2}}+\frac{1}{r^{3}}\left(\frac{Q^{2}f'(0)}{4f(0)^{2}}-\phi'_{3}(t)\right)+\frac{3}{2\Lambda r^{4}}\phi''_{3}(t)+O(r^{-5}),
\end{eqnarray}
where $f'(0)=\frac{df(\phi)}{d\phi}|_{\phi=0}$, and $\phi'_{3}(t)=\frac{d\phi_{3}(t)}{dt}$.
The expansions are   totally determined by parameters  $\Lambda$, $M$ and $Q$ and a function $\phi_{3}(t)$. % which is the coefficient of $r^{-3}$.
Parameters $M$ and $Q$ could be interpreted as the Arnowitt-Deser-Misner (ADM) mass and charge of the spacetime, respectively.
%Parameter $c$ will be set to zero $c=0$ for simplicity in this work.
Note that we have required   $\zeta-r=0$   as $r\to\infty$ by fixing the residual radial reparameterization freedom \cite{Chesler:2013lia}.

Given parameters $\Lambda$, $M$, $Q$ and the scalar perturbation $\phi(t=0,r)$ on the initial time slice, we  integrate (\ref{eq:alphar}-\ref{eq:Pr}) radially inwards to get the  initial $\alpha,S,\zeta,P$ subjected to the asymptotic boundary solutions.
Then we can work out  $\phi$  on the next time slice using eq.(\ref{eq:Pt}).
 Repeating the procedure iteratively, we get $\alpha$, $\zeta$ $S$ and $P$ on all time slices.
Equations (\ref{eq:St},\ref{eq:S}) are redundant and can be used to check the accuracy of the code.

Variables $\alpha,S,\zeta$  diverge at infinity, new variables $\sigma\equiv \zeta/r,a\equiv \alpha/r^{2},s\equiv S/r^{2},p\equiv rP$ are introduced to implement the numerical method.
In asymptotic AdS spacetime, the scalar waves can propagate to the infinity at finite coordinate time and be bounced back to the bulk. The infinity must be included in the computational
domain. We  use a new coordinate $z=\frac{r}{r+1}$ to compactify the radial domain to $(z_{0},1)$.
Here $z_{0}$ corresponds to some radius $r_{0}$ which is close to the black hole apparent horizon $r_{a}$ from inside.
In time direction the system is evolved with fourth order Runge-Kutta scheme, the radial direction is discretized in a uniform grid.
Equations (\ref{eq:Sr}) is discretized with second order finite difference while (\ref{eq:alphar},\ref{eq:zetar},\ref{eq:Pr}) with fourth order finite difference, as indicated in \cite{Bosch:2016vcp}.
At the boundary some terms in the equations appear as the $0/0$-type, the l'Hôpital's rule is implemented to reduce the instabilities. % from small denominators.
The Kreiss-Oliger dissipation is employed to stabilize the numerical evolution as well.
%The equations in $z$ coordinate for numerical calculation are as follows.
%\begin{eqnarray}
%0&=& \partial_{t}\phi+\frac{1}{2}az^{2}\partial_{z}\phi+\frac{p(z-1)}{z},\label{eq:phitz}\\
%0&=& \partial_{z}^{2}\sigma+\frac{2}{z}\partial_{z}\sigma+\sigma(\partial_{z}\phi){}^{2},\label{eq:sigmaz}\\
%0&=& \partial_{z}s+\frac{s\partial_{z}\sigma}{\sigma}-\frac{(z-1)^{3}Q^{2}}{2z^{5}\sigma^{3}f(\phi)}+\frac{(z-1)}{2\sigma z^{3}}-\frac{6s+\Lambda\sigma}{2(z-1)z}\label{eq:sz}\\
%0&=& \partial_{z}p+\frac{p\partial_{z}\sigma}{\sigma}+\frac{z^{2}}{(z-1)^{2}}\frac{s\partial_{z}\phi}{\sigma}-\frac{(z-1)Q^{2}}{4z^{3}\sigma^{4}f(\phi)^{2}}\frac{df(\phi)}{d\phi}\label{eq:pz}\\
%0&=& \partial_{z}^{2}a+\frac{2\left((z-2)\partial_{z}a+2s\frac{\partial_{z}\sigma}{\sigma^{2}}\right)}{(z-1)z}-\frac{4p(z-1)\partial_{z}\phi}{z^{3}}\nonumber\\&&+\frac{2a\sigma-4s}{\sigma(z-1)^{2}z^{2}}+\frac{2}{\sigma^{2}z^{4}}-\frac{4Q^{2}(z-1)^{2}}{f(\phi)\sigma^{4}z^{6}}\label{eq:az}
%\end{eqnarray}
%
%The asymptotic solutions in $z$ coordinate turn to be
%\begin{align}
%\phi(z) & =c+\phi_{3}(t)(1-z)^{3}+O((1-z)^{4})\\
%\sigma(z) & =1-\frac{3\phi_{3}^{2}(t)}{10}(1-z)^{6}+O((1-z)^{7})\\
%a(z) & =-\frac{\Lambda}{3}+(1-z)^{2}+\left(\frac{Q^{2}}{f(c)}-3\right)(1-z)^{4}+O((1-z)^{5})\\
%s(z) & =-\frac{\Lambda}{6}+\frac{(1-z)^{2}}{2}+\frac{(1-z)^{4}}{2}\left(\frac{Q^{2}}{f(c)}-3\right)+O((1-z)^{5})\\
%p(z) & =\frac{\Lambda\phi_{3}(t)}{2}(1-z)+O((1-z)^{2})
%\end{align}

	We also provide a numerical exploration for phase structure. Locating the critical point where the scalarisation occurs is to find a static normalizable mode of the scalar field. When the scalarisation starts to occur, $\phi$ is a static normalizable mode that can be dealt with by perturbation. Following the methods provided in \cite{Horowitz:2013jaa}, we turn the problem of locating the critical point of the scalarisation to the problem of solving the eigenvalue problem of $b$. The perturbation equation of \eqref{scalarequation} is,
	\begin{equation}\label{eq:varyphi}
	  \nabla^\mu \nabla_\mu \delta\phi = -\frac{b}{2} F_{\mu\nu}F^{\mu\nu} \delta\phi.
	\end{equation}
	In order to solve the eigenvalue problem \eqref{eq:varyphi}, we work in coordinate $z\equiv r_h/r$, and discretize the $z\in (0,1)$ such that \eqref{eq:varyphi} becomes a linear algebra problem. We require $\phi$ to vanish at the boundary and to be regular at the horizon. Note that in \cite{Guo:2020sdu} the phase structure of the scalarisation has been investigated through the stability with time-domain analysis. Our treatment for the phase structure is easier and more direct, and our results match with those in \cite{Guo:2020sdu}. Also, the critical points in our later dynamical evolution also match the results from the eigenvalue problem analysis.

\section{Numerical results}

%This section summarize the numerical results obtained by the evolution scheme of the last section.
%In Section. \ref{initial}, we describe the initial profile of scalar field and the value of free parameters determining this model.
%Section.\ref{decay} gives the results on the cases evolving into a final state which is a RN-AdS black hole.
%The comparison between the analytical and numerical results of metric are provided as a way of checking the accuracy of our code.
%In Section.\ref{con} we summarize the general properties of the evolution of spacetime whose final state employs non-trivial scalar.
%The effects of charge $Q$ and coupling parameter $b$ on the final states are investigated.
%With the help of discrete Fourier transforms we also analyze the evolution of the scalar modes with different frequencies.

\subsection{Initial data}\label{initial}
 We fix $M=1$ in the numerical calculations to investigate how  charge $Q$, coupling constant $b$ and the cosmological constant $\Lambda$ affect the evolution of the gravitational system.
For the initial profiles of the scalar field, we use two different families:
\begin{equation}
\phi(t=0,r)=\kappa~e^{-\frac{\left(r-4r_{h}\right)^{2}}{w^{2}}}
\end{equation}
and
\begin{equation}
\phi(t=0,r)=\begin{cases}
\left(\frac{1}{r}-\frac{1}{r_{1}}\right)^{3}\left(\frac{1}{r}-\frac{1}{r_{2}}\right)^{3}\frac{\kappa_{1}+\kappa_{2}\sin\frac{10}{r}}{r^{2}}, & r_{1}<r<r_{2},\\
0 & r\le r_{1}\text{ or }r\ge r_{2}.
\end{cases}
\end{equation}
Here $\kappa<10^{-9}$ and width $w=1.8r_{h}$ where $r_{h}$ is the horizon of the corresponding RN-AdS black hole with the same
$M,Q$ and $\Lambda$. $\{r_{1},r_{2}\}=\{2r_{h},3r_{h}\}$ and $\kappa_{1},\kappa_{2}$
are of order $10^{-2}$. As a consequence, the initial scalar field
is of order $10^{-10}$ so that the scalar field is initially negligible
compared to the black hole.

%\subsection{RN-AdS black hole as a solution to the model}\label{decay}
When the scalar field $\phi$ employs the trivial configuration $\phi_3=0$, one could solve the equations analytically. The solution  is nothing but the RN-AdS black hole, in which the metric functions are given as $\alpha(r)=r^2-2M/r+Q^2/r^2,\zeta=r.$

%We also could choose the appropriate values of parameters of $Q$ and $b$, let the gravitational system evolve into a RN-AdS black hole with trivial scalar configuration. We solve such a scheme, and compare the metric of final black hole with the analytical results as a check of the accuracy of our code.
%In fig.(\ref{}) we set the $Q=1$ and $b=1$,$Q=1$, the red lines are analytical results while blue are obtained numerically. The difference between them are showed in yellow which are at order $10^{-10}$.

\subsection{Evolving into scalar hairy black hole}\label{con}

What is more interesting is the spacetime with a non-trivial scalar field configuration that will be discussed in details in following subsections.

\subsubsection{Effects of coupling parameter $b$ on the black hole spontaneous scalarisation}

\begin{figure*}
\begin{centering}
{\footnotesize{}}%
\begin{tabular}{cc}
{\footnotesize{}\includegraphics[width=0.45\textwidth]{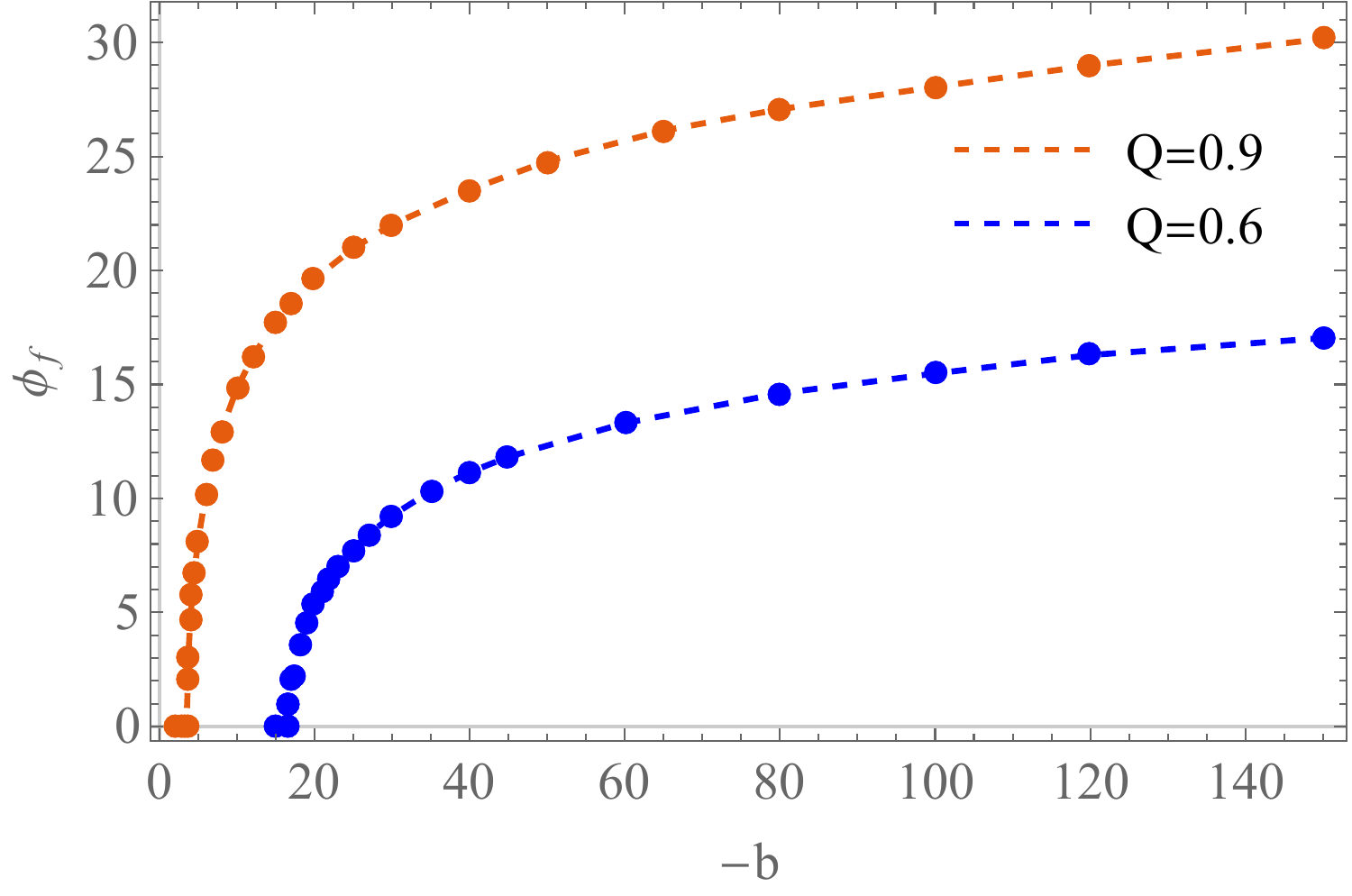}}\tabularnewline
\end{tabular}{\footnotesize\par}
\par\end{centering}
{\footnotesize{}\caption{\label{fig:Q96b3020phi3fl}The final value $\phi_{f}$ of $\phi_{3}(t)$
for various $b$ when $Q=0.9$ and $0.6$. Here $\Lambda=-0.03$.}
}{\footnotesize\par}
\end{figure*}

In this subsection, we fix  the cosmological constant $\Lambda=-0.03$, choose the black hole charge $Q=0.6,09$ and change $b$ to study the effects of the coupling parameter $b$ on dynamics of the black hole spontaneous scalarisation. Since $\phi_3$ can be viewed as an indicator of the nontrivial scalar hair in some sense, we first show the final value $\phi_{f}$ of $\phi_{3}$ when the system settles down for various $b$  in Fig.\ref{fig:Q96b3020phi3fl}.
For $Q=0.9$, when  $-b$ is smaller than a critical value  $-b_\ast=3.5$, the final value $\phi_f$ vanishes and the gravitational system evolves into a RN-AdS black hole. When charge $Q=0.6$, the critical value $b_{\ast}\simeq-16.5$.
When $-b\gtrsim-b_\ast$, the final value $\phi_f$ no longer vanishes. A static non-trivial scalar configuration forms outside the black hole.
The hairy black hole solution exists only when $b$ is negative enough. We checked that the domain of existence for scalarized black hole is consistent with that of \cite{Guo:2020sdu}. Near the critical
value $b_\ast$ where spontaneous scalarisation is triggered, $\phi_f$ changes unsmoothly. This is similar to the case  in asymptotic flat spacetime \cite{Herdeiro:2018wub}.
{\footnotesize{}}
{\footnotesize\par}

\begin{figure*}
\begin{centering}
{\footnotesize{}}%
\begin{tabular}{cc}
{\footnotesize{}\includegraphics[width=0.45\textwidth]{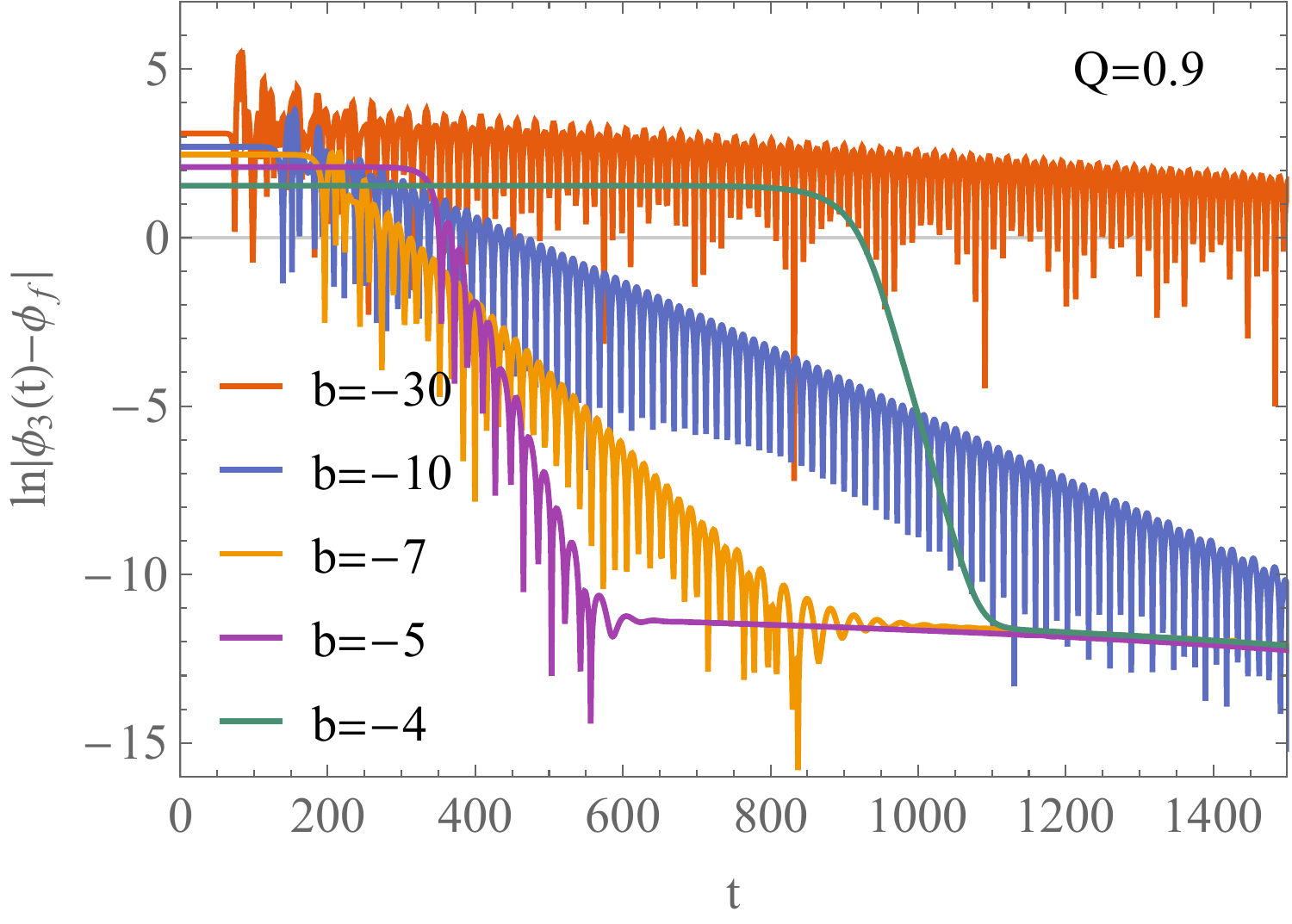}} & {\footnotesize{}\includegraphics[width=0.45\textwidth]{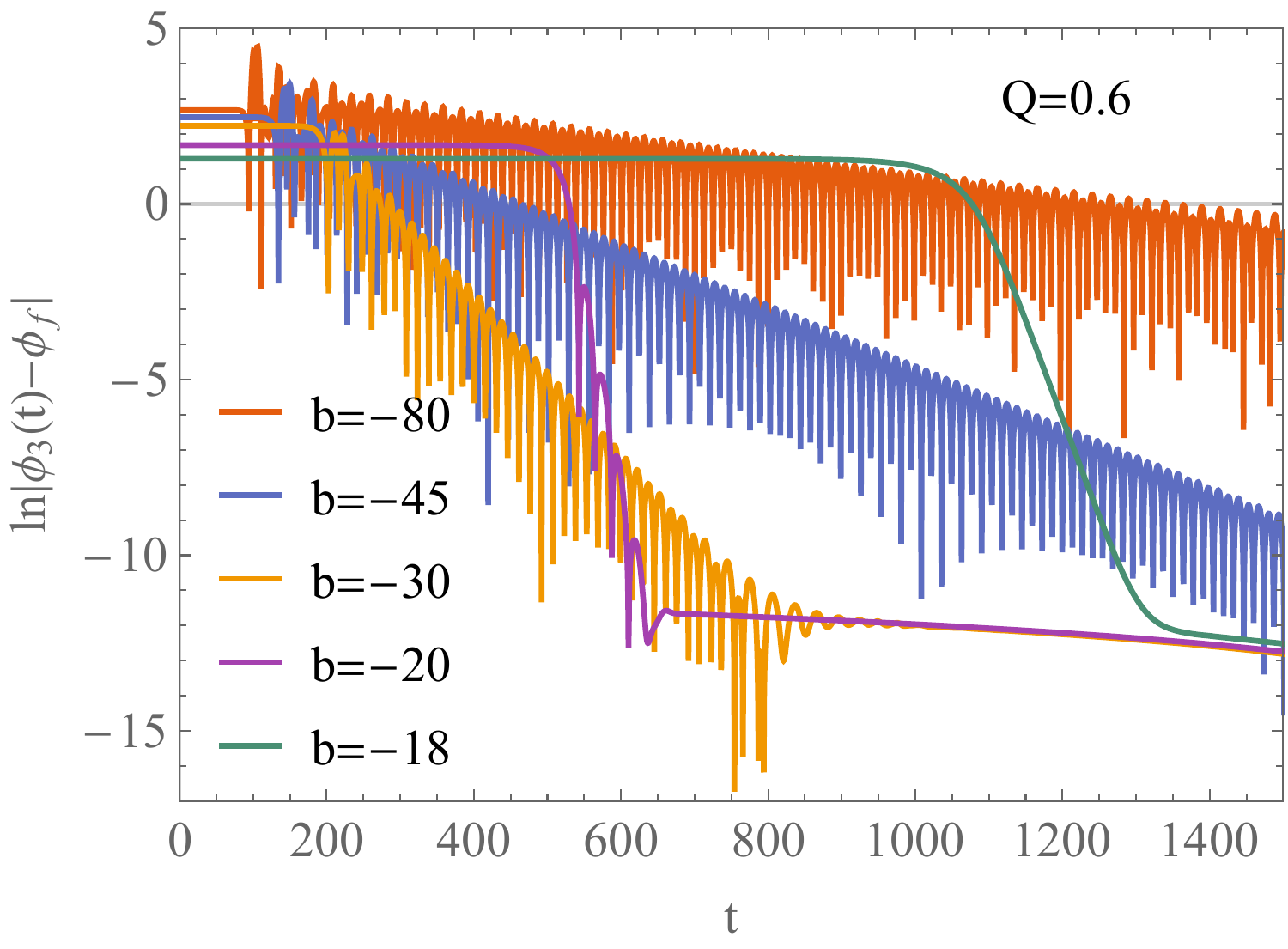}}\tabularnewline
\end{tabular}{\footnotesize\par}
\par\end{centering}
{\footnotesize{}\caption{\label{fig:Q96phi3}{\small{} }The evolution of $\ln|\phi_{3}(t)-\phi_{f}|$
for various $b$ when $Q=0.9$ (left) and $Q=0.6$ (right). Here  $\Lambda=-0.03$.}
}{\footnotesize\par}
\end{figure*}

Now we study the evolution of $\phi_3$.
It oscillates with damping amplitude and converges to the final value $\phi_{f}$. We plot $\ln|\phi_{3}(t)-\phi_{f}|$ in Fig.\ref{fig:Q96phi3} for various $b$.
The general features resemble the behavior of quasinormal modes which can be divided into three stages.
At early times, $\phi_{3}$ changes little.
Then it damps exponentially.
At late times, it converges to $\phi_{f}$ by a power-law. When $b$ is close to the critical value $b_{\ast}$, the scalar decay to $\phi_{f}$ without oscillating.
%Near the critical value of $b$ where spontaneous scalarisation is triggered, the scalar hair changes unsmoothly near the $b_{\ast}$.

\begin{figure*}
\begin{centering}
{\footnotesize{}}%
\begin{tabular}{ccc}
{\footnotesize{}\includegraphics[width=0.33\textwidth]{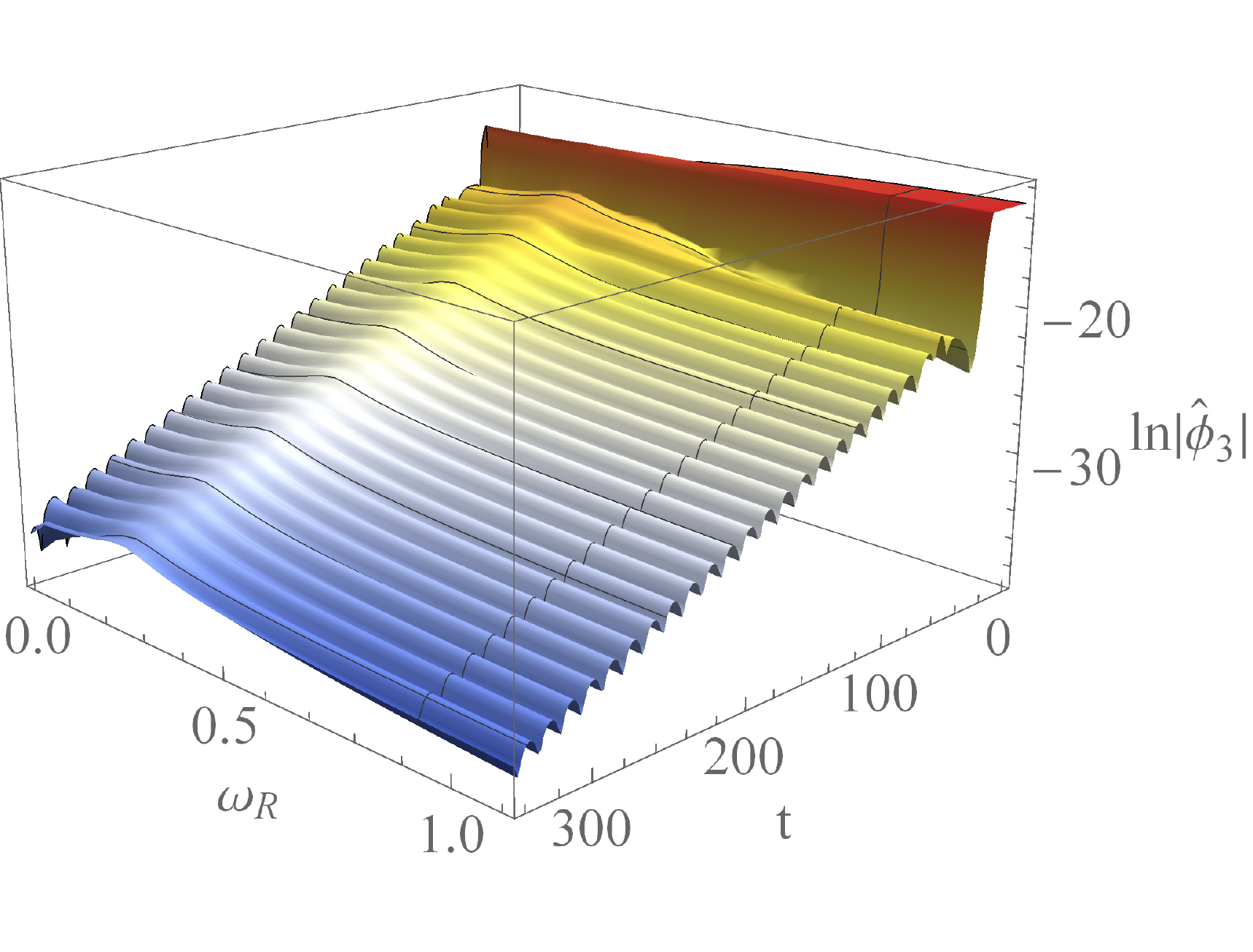}} &
{\footnotesize{}\includegraphics[width=0.33\textwidth]{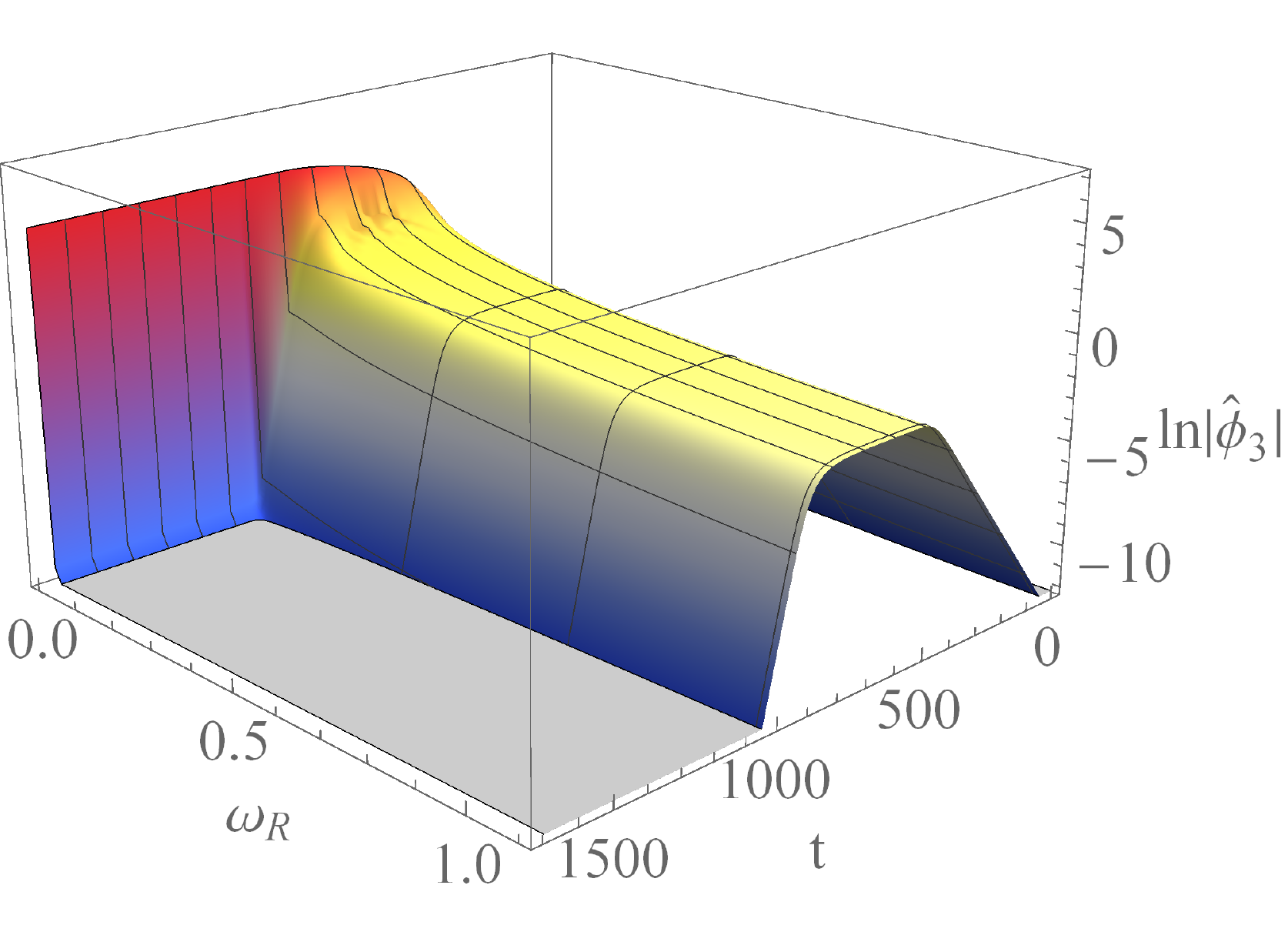}} & {\footnotesize{}\includegraphics[width=0.33\textwidth]{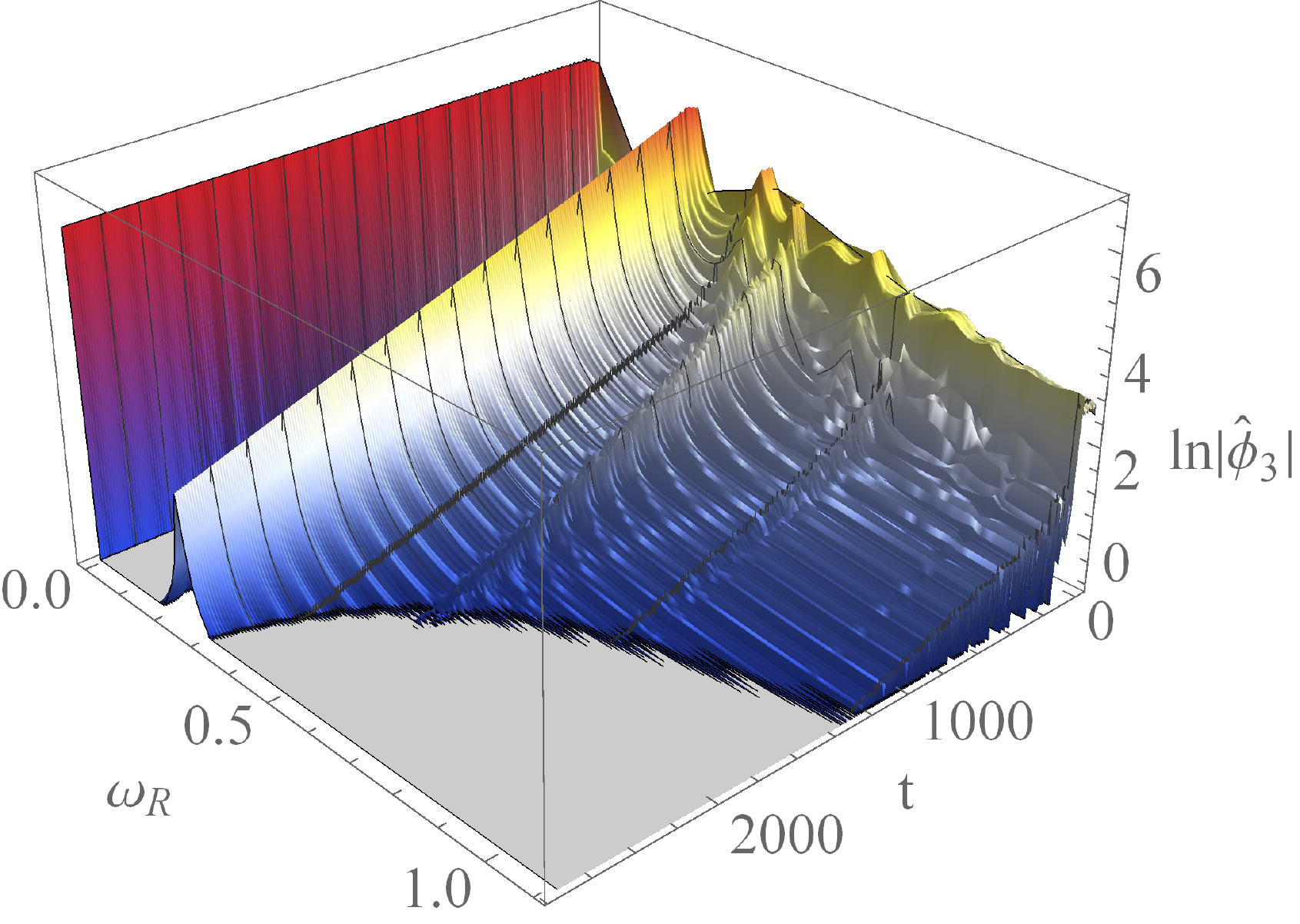}}\tabularnewline
\end{tabular}{\footnotesize\par}
\par\end{centering}
{\footnotesize{}\caption{\label{fig:Q9phi3Fourier}{\small{} }The evolution of the logarithm
of the amplitude of the discrete Fourier transformation of $\phi_{3}(t)$
for $b=-2$ (left), $b=-4$ (middle) and $b=-30$ (right) when $Q=0.9$ and $\Lambda=-0.03$. }
}{\footnotesize\par}
\end{figure*}

{\footnotesize{}}
To investigate the evolution  of each frequency component of the scalar field, we adopt the similar procedure in  ref.\cite{Bosch:2016vcp}.  Partitioning the time axis into intervals, and performing the discrete Fourier transformation on these intervals, we get the evolution of the amplitudes of each frequency component.
Fig.\ref{fig:Q9phi3Fourier} shows some spectrograms of the evolution.
In the left panel where $b=-2$, all the modes of the  scalar decay exponentially and finally be  absorbed by the black hole and leads to a RN-AdS black hole solution.  In the middle panel where $b=-4$,   all the modes of the scalar grows exponentially at early times.
At late times,  the zero mode with $\omega_{R}=0$  almost keep constant while the other modes decay exponentially.
The right panel of Fig.\ref{fig:Q9phi3Fourier} depicts the evolution of the system with $b=-30$. The evolution of amplitudes becomes irregular. The components with higher frequency decay directly, while the modes with smaller frequency grow rapidly at first and then decay. The zero mode keeps constant at late times.
Note that a non-vanishing zero mode at late times indicates the non-vanishing scalar hair.

\begin{figure*}
\begin{centering}
{\footnotesize{}}%
\begin{tabular}{cc}
{\footnotesize{}\includegraphics[width=0.42\textwidth]{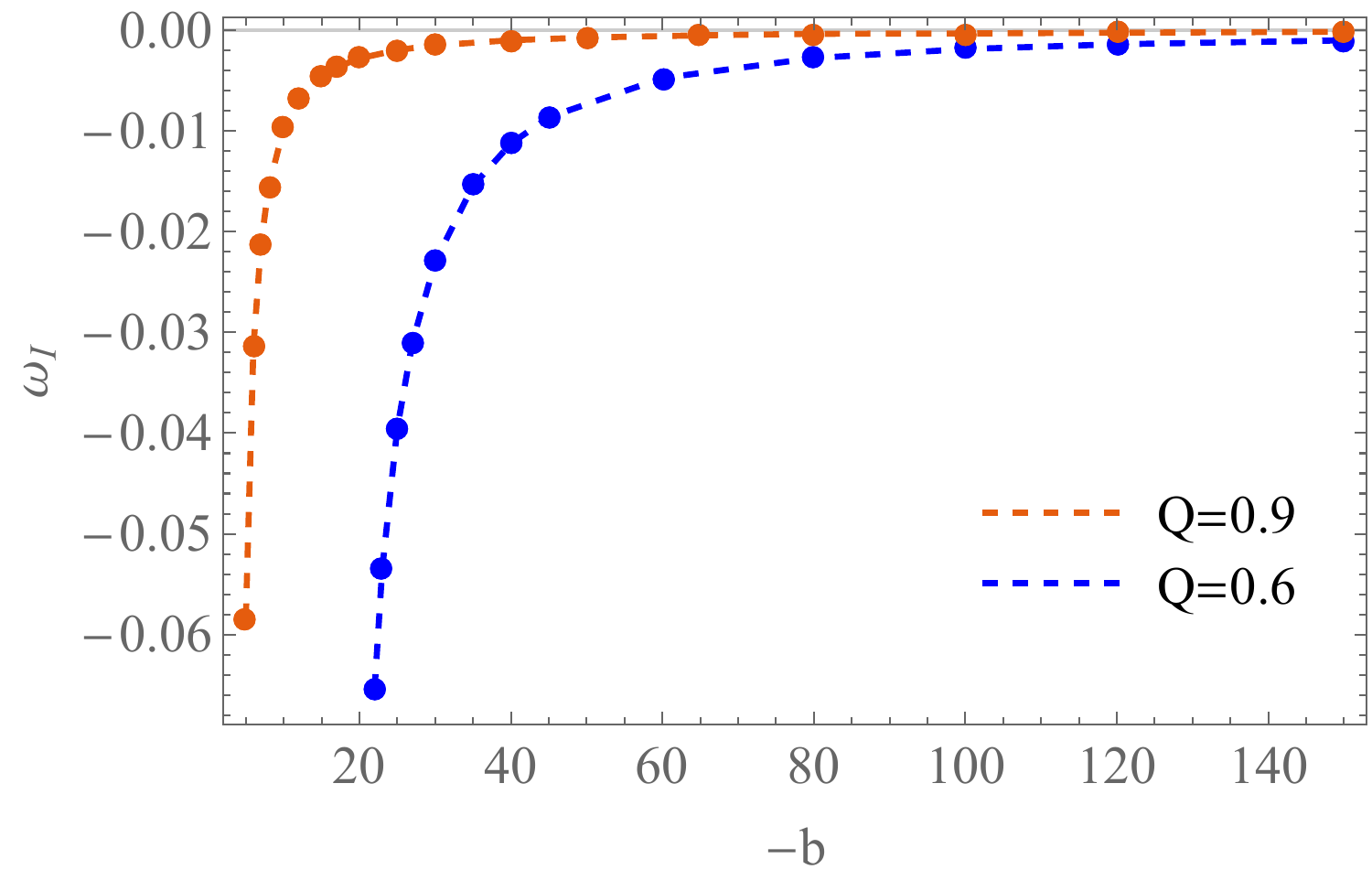}} & {\footnotesize{}\includegraphics[width=0.42\textwidth]{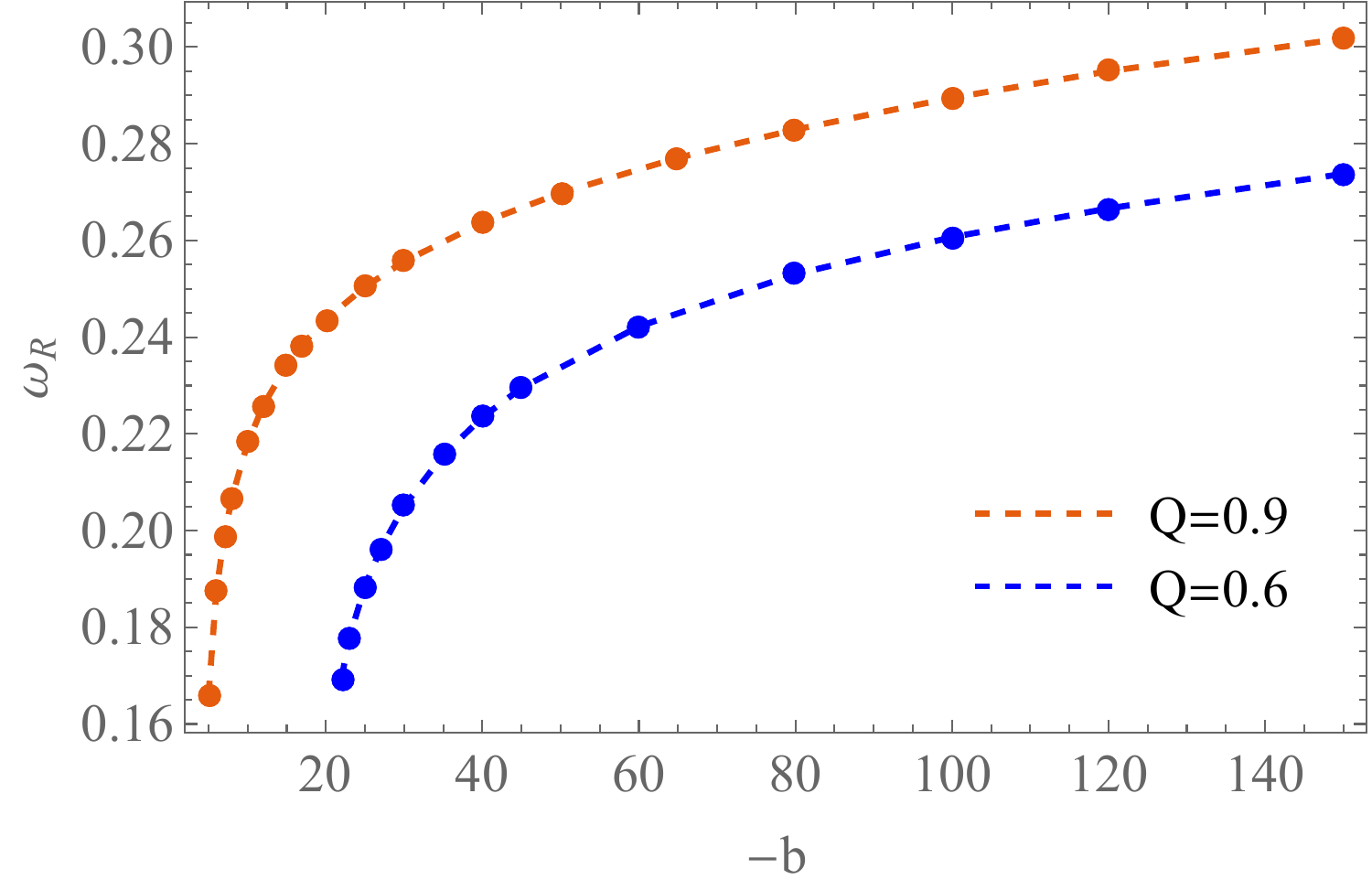}}\tabularnewline
\end{tabular}{\footnotesize\par}
\par\end{centering}
{\footnotesize{}\caption{\label{fig:Q96wIR}{\small{} }The complex frequencies of the dominant damping modes of $\phi_{3}$
for various $b$ when $Q=0.9$ and $0.6$. Left for the imaginary
part $\omega_I$ and right for the real part $\omega_R$ of the corresponding frequencies. Here  $\Lambda=-0.03$.}
}{\footnotesize\par}
\end{figure*}

{\footnotesize{}}
{\footnotesize\par}
The damping rate of the dominant damping modes can be extracted by fitting the slope of the corresponding  logarithm of the amplitude in Fig.\ref{fig:Q9phi3Fourier}. The more efficient way is by using Prony method which can work out the complex frequency directly \cite{Berti:2007dg}.
We checked that these two methods give consistent results when they are both valid.
The results are shown in Fig.\ref{fig:Q96wIR}.
As $-b$ increases, the imaginary part $\omega_{I}$ of the dominant damping mode tends to zero. The real part $\omega_R$ increases monotonically. This implies the system oscillates faster and  needs more time to settle down as  $-b$ increases.

{\footnotesize{}}
{\footnotesize\par}

\begin{figure*}
\begin{centering}
{\footnotesize{}}%
\begin{tabular}{cc}
{\footnotesize{}\includegraphics[width=0.42\textwidth]{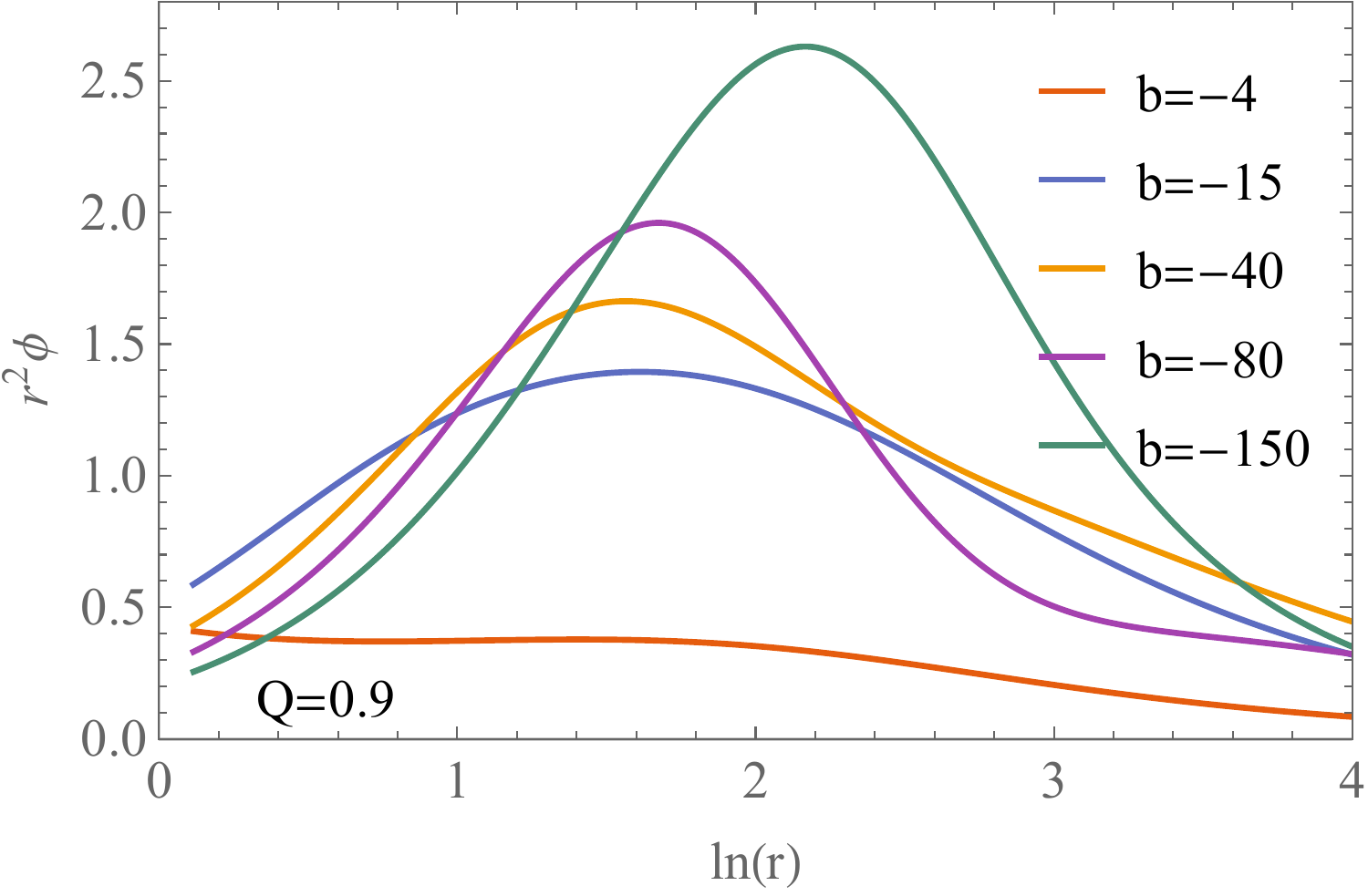}} & {\footnotesize{}\includegraphics[width=0.42\textwidth]{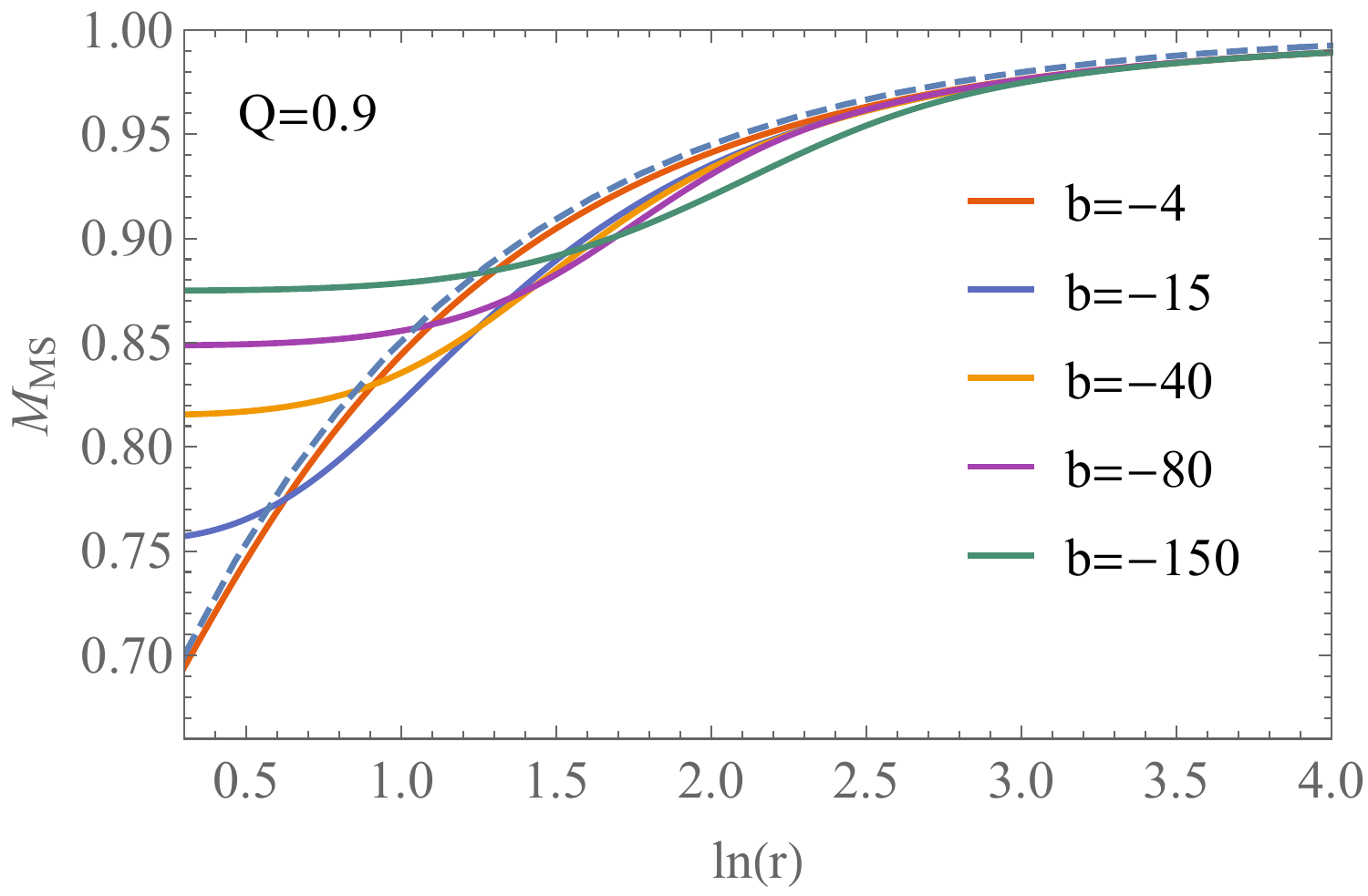}}\tabularnewline
{\footnotesize{}\includegraphics[width=0.42\textwidth]{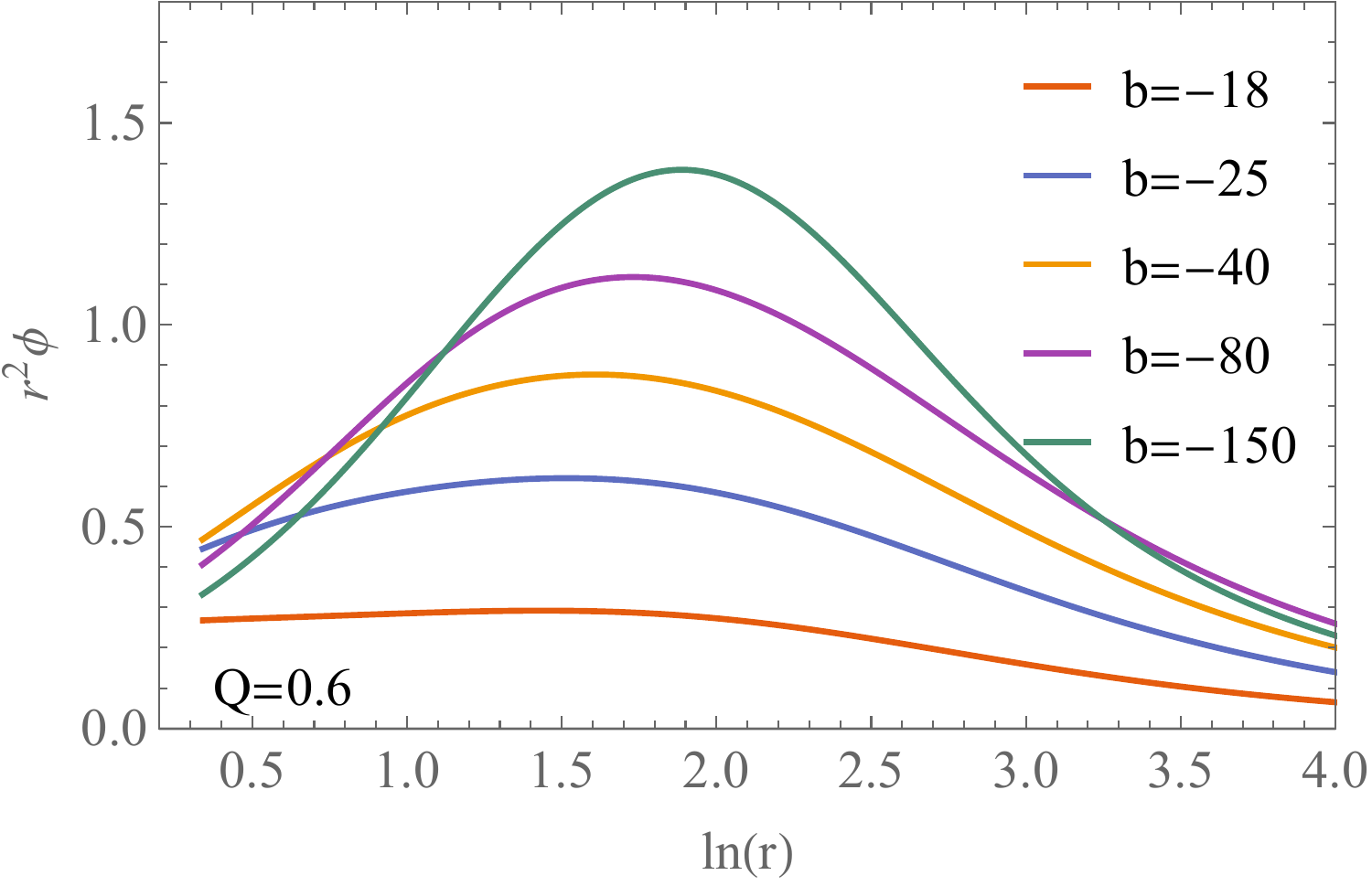}} & {\footnotesize{}\includegraphics[width=0.42\textwidth]{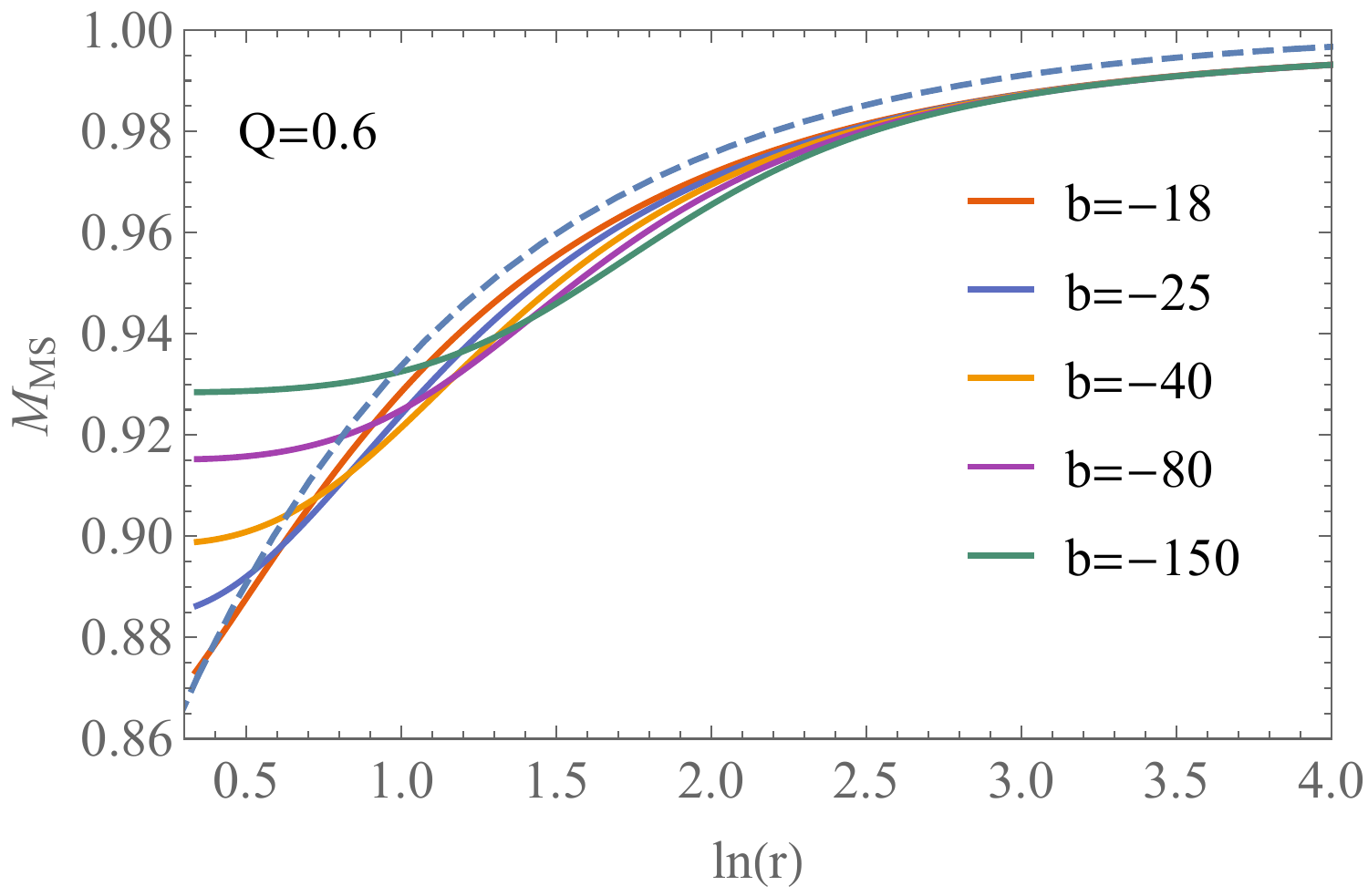}}\tabularnewline
\end{tabular}{\footnotesize\par}
\par\end{centering}
{\footnotesize{}\caption{\label{fig:Q96scalarProfile}{\small{} }The rescaled scalar profile
$r^{2}\phi$ (left) and Misner-Sharp mass (right) of the final hairy
black holes when $Q=0.9$ (upper) and $0.6$ (lower) for various $b$.
The dashed line in the right panels correspond to the Misner-Sharp
mass of the RN-AdS black hole with $Q=0.9$ and $Q=0.6$, respectively. Here $\Lambda=-0.03$.}
}{\footnotesize\par}
\end{figure*}

We show the rescaled final scalar profile of the hairy black hole for various $b$ when $Q=0.9$ and $0.6$ in the left of Fig.\ref{fig:Q96scalarProfile}, from which we can see the final scalar configuration  becomes heavier as $-b$ increases.
In the right of Fig.\ref{fig:Q96scalarProfile} we show the corresponding final rescaled Misner-Sharp mass $M_{MS}=m/4\pi$.
Here the generalized Misner-Sharp quasi-local mass is defined by \cite{Maeda:2012fr}
\begin{equation}
m =2\pi\zeta\left(-\frac{\Lambda}{3}\zeta^{2}+1-2S\partial_{r}\zeta\right).
\end{equation}
As $r\to\infty,$ the Misner-Sharp mass $M_{MS}$ tends to the ADM mass $M$.
The Misner-Sharp mass at the horizon increases as $-b$ increases.
For larger $-b$, the Misner-Sharp mass is almost a constant
near the horizon and increases with $r$ when $r$ is large. This
is consistent with the left of Fig.\ref{fig:Q96scalarProfile} where
the the scalar field localized further away from the black hole as $-b$ increases. Note that
unlike $\phi_{f}$ at the infinity which increases as $-b$ monotonously,
the value of the scalar on the apparent horizon is not monotonous.
This is similar to the case in asymptotic flat spacetime \cite{Herdeiro:2018wub}.
{\footnotesize{}}
{\footnotesize\par}

\begin{figure*}
\begin{centering}
{\footnotesize{}}%
\begin{tabular}{c} {\footnotesize{}\includegraphics[width=0.47\textwidth]{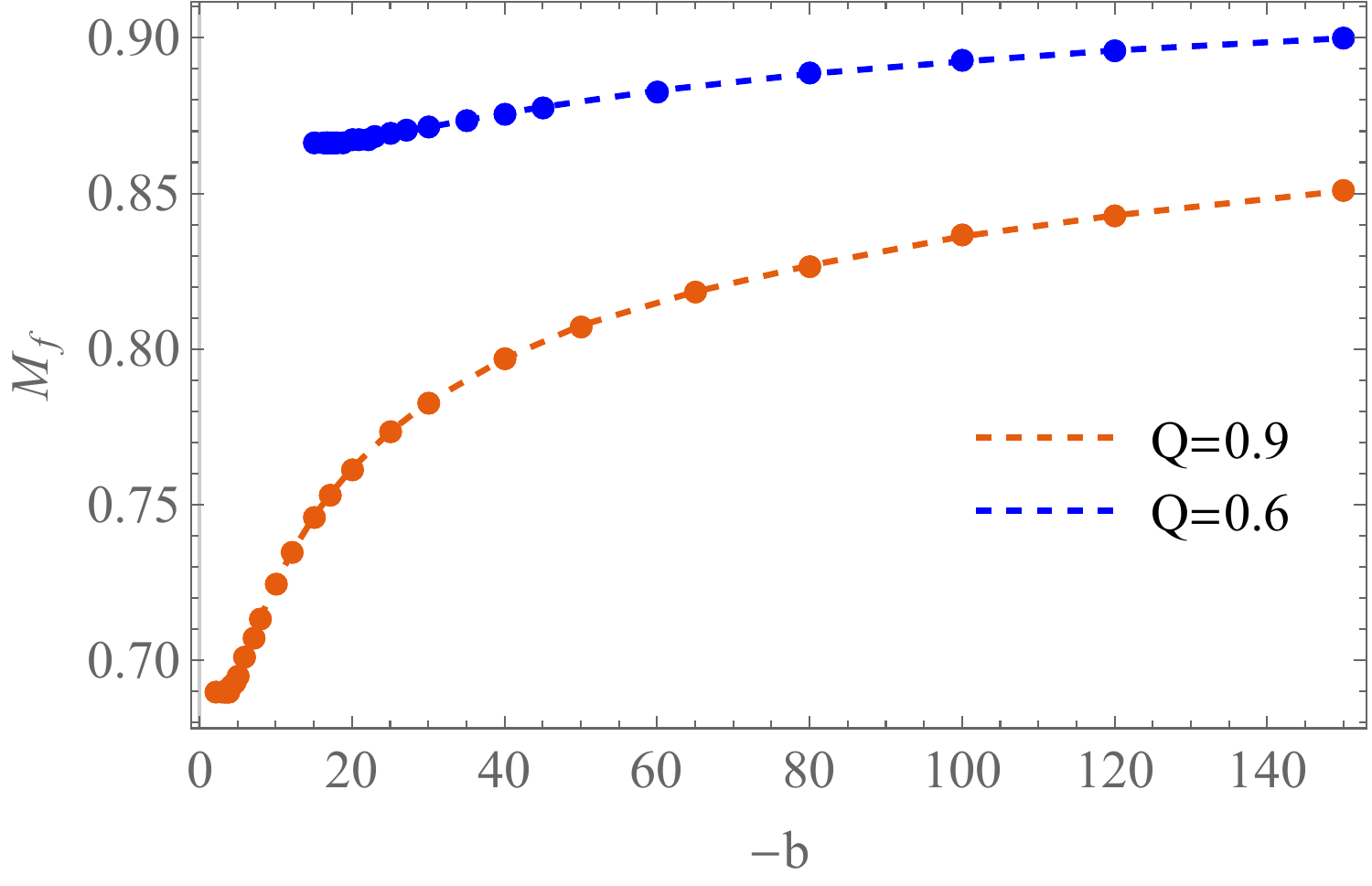}}\tabularnewline
\end{tabular}{\footnotesize\par}
\par\end{centering}
{\footnotesize{}\caption{\label{fig:Q96irrMf}{\small{} }The final value of irreducible mass
$M_{f}$ of the black hole for various $b$ when $Q=0.9$ and $0.6$.}
}{\footnotesize\par}
\end{figure*}

Now we study the evolution of the irreducible mass of the black hole which is defined as
%\begin{equation}
%M_{0}(t)\equiv\sqrt{\frac{A_{AH}}{4\pi}}=\zeta(r_{a},t).
%\end{equation}
$M_{0}(t)\equiv\sqrt{\frac{A_{AH}}{4\pi}}=\zeta(r_{a},t).$
Here $A_{AH}$ is the area. $r_a$ is the location of apparent horizon which satisfies the equation
\begin{equation}
0=g^{AB}\partial_{A}\zeta\partial_{B}\zeta=2S\partial_{r}\zeta.
\end{equation}
The final value $M_{f}$ of $M_0(t)$ is depicted in Fig.\ref{fig:Q96irrMf}.
Starting from the same initial value $M_{i}$, a stronger coupling corresponding to a more negative $b$ leads to a larger final irreducible mass.
Unlike the case of the scalar hair $\phi_{f}$, the final irreducible mass changes smoothly near $b_{\ast}$.
%When $b$ is close to the critical value $b_{\ast}$ where the spontaneous scalarisation just occurs, the time $M_{0}(t)$ starts to grow tends to infinity.

{\footnotesize\par}

\begin{figure*}
\begin{centering}
{\footnotesize{}}%
\begin{tabular}{cc}
{\footnotesize{}\includegraphics[width=0.42\textwidth]{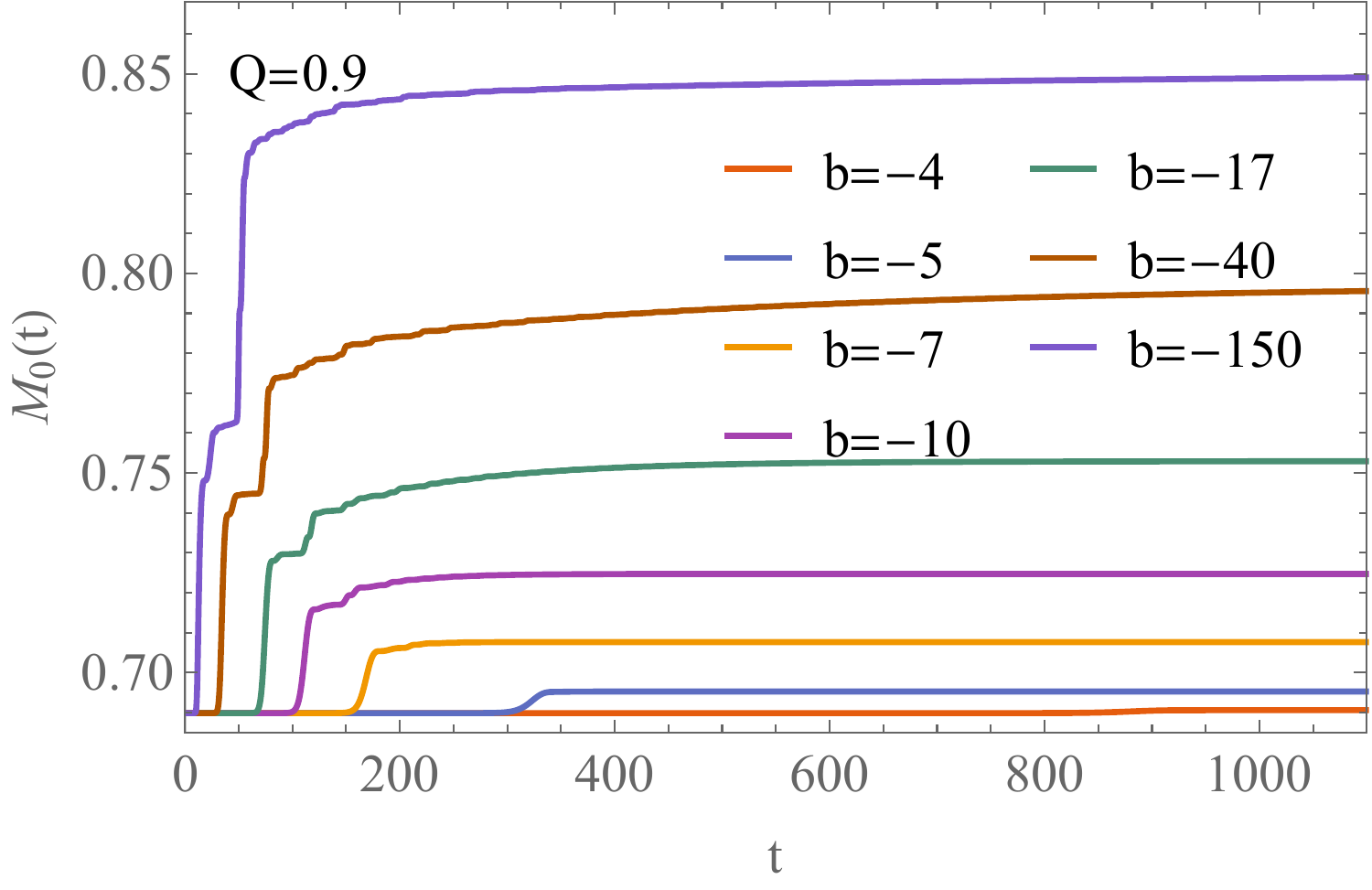}} & {\footnotesize{}\includegraphics[width=0.42\textwidth]{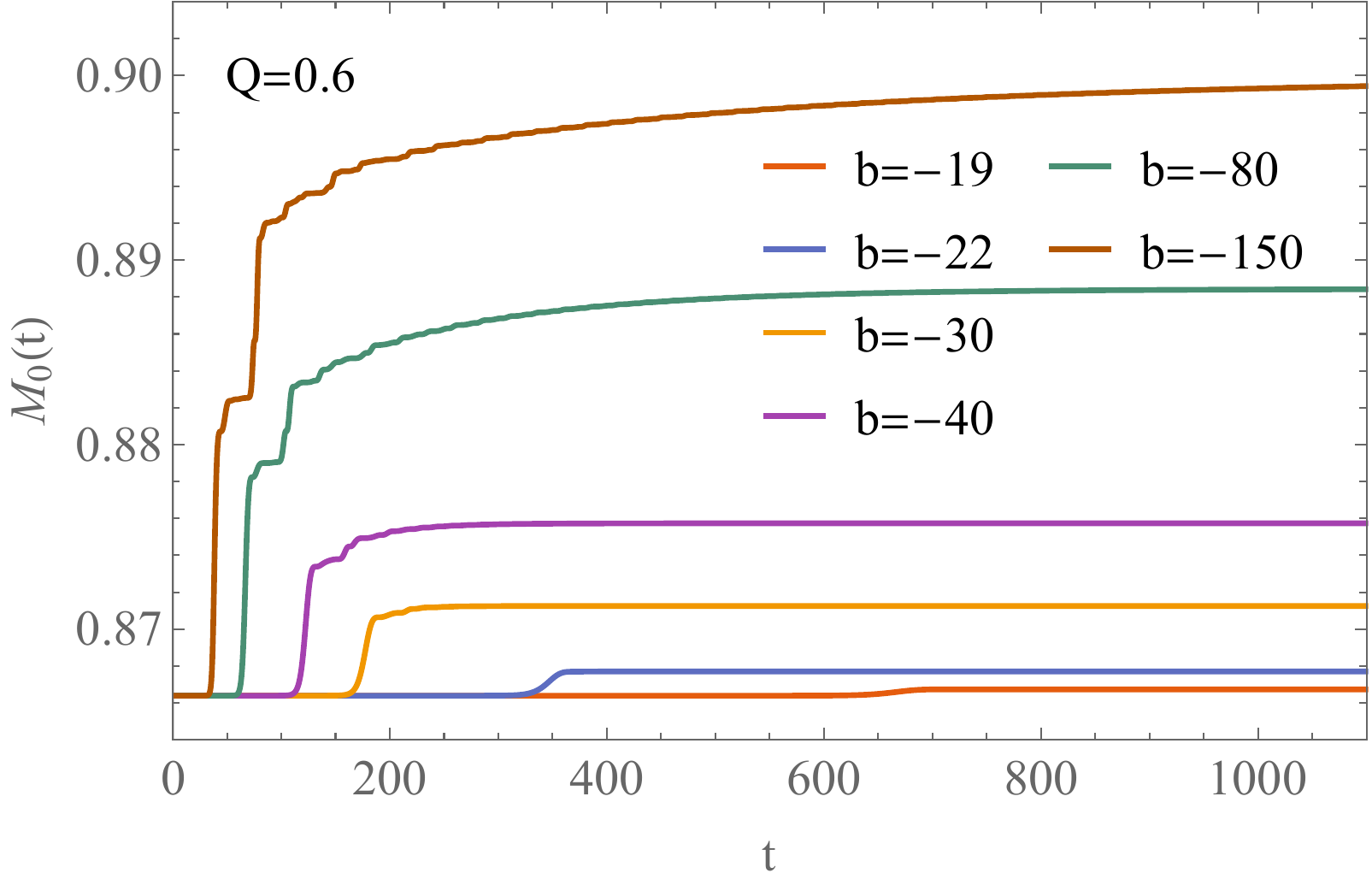}}\tabularnewline
\end{tabular}{\footnotesize\par}
\par\end{centering}
{\footnotesize{}\caption{\label{fig:Q96irrM}{\small{} }The evolution of the irreducible mass $M_{0}(t)$
of the black hole for various $b$ when $Q=0.9$ and $0.6$. Here $\Lambda=-0.03$.}
}{\footnotesize\par}
\end{figure*}

In Fig.\ref{fig:Q96irrM} we show the evolution of $M_{0}$ for various $b$ when $Q=0.9$ and $0.6$. The irreducible mass never decreases with time. We checked that  both the null energy condition and weak energy condition are satisfied in the evolution.
{\footnotesize{}}

{\footnotesize{}}
{\footnotesize\par}

\begin{figure*}
\begin{centering}
{\footnotesize{}}%
\begin{tabular}{cc}
{\footnotesize{}\includegraphics[width=0.42\textwidth]{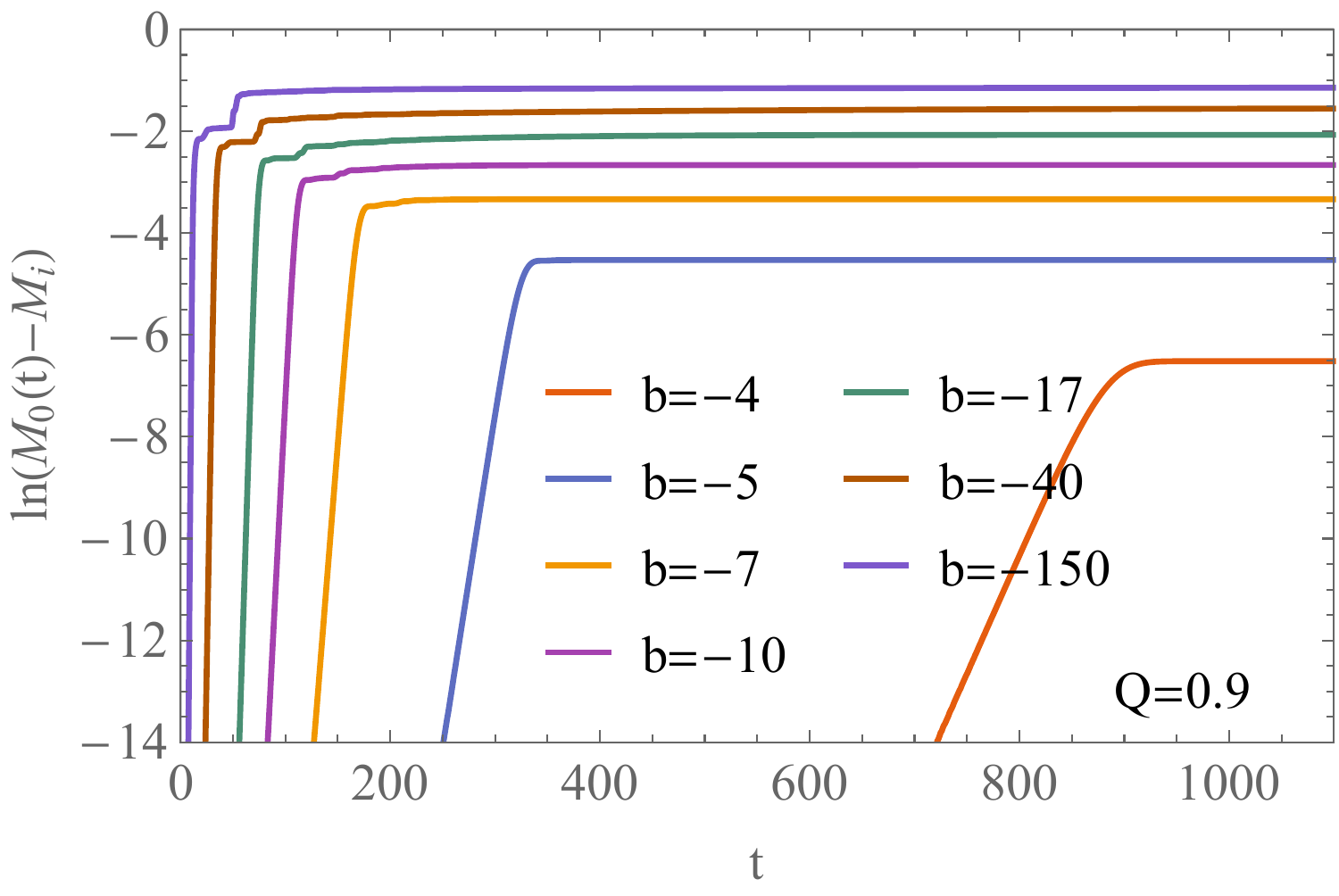}} & {\footnotesize{}\includegraphics[width=0.42\textwidth]{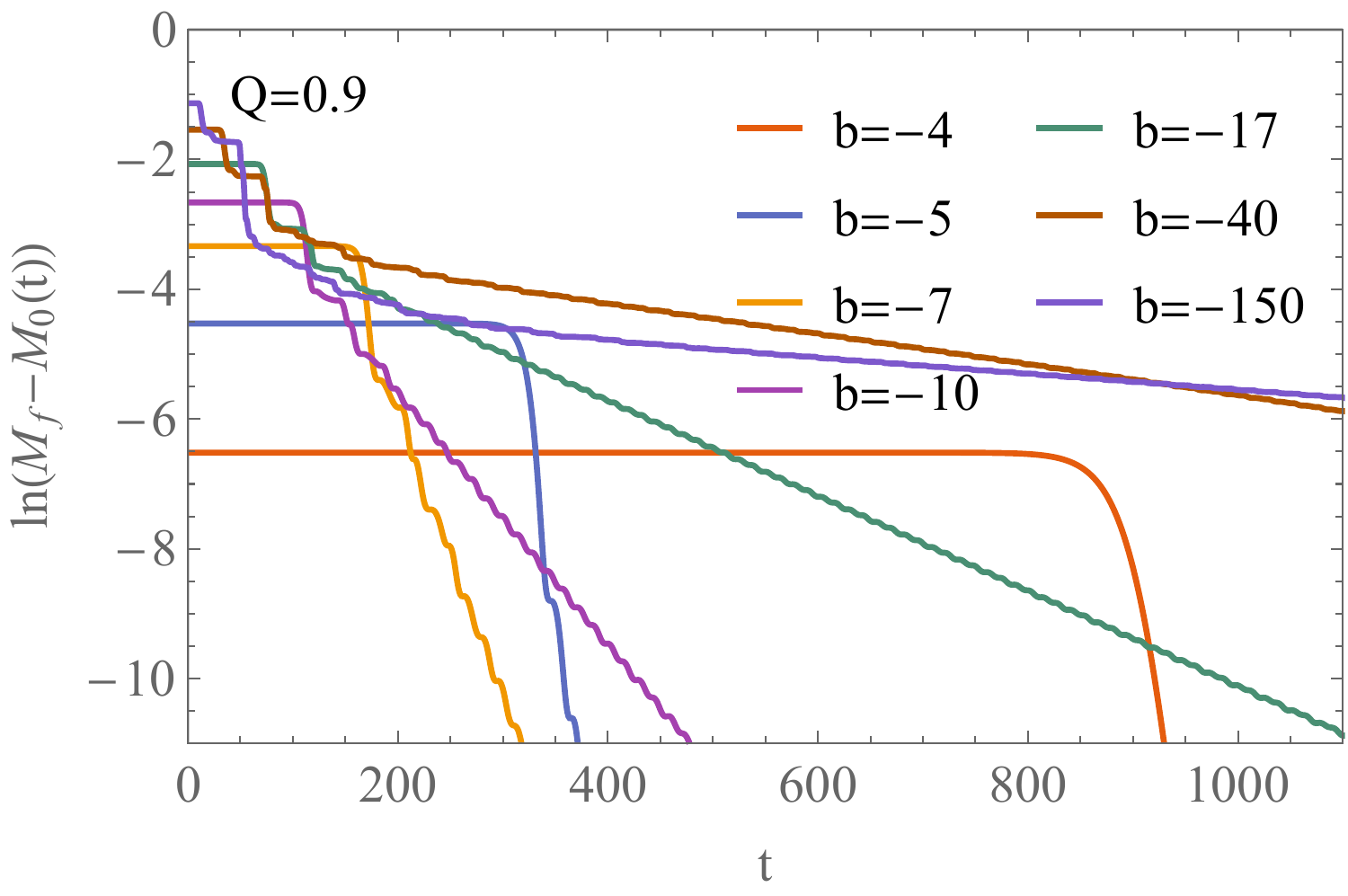}}\tabularnewline
{\footnotesize{}\includegraphics[width=0.42\textwidth]{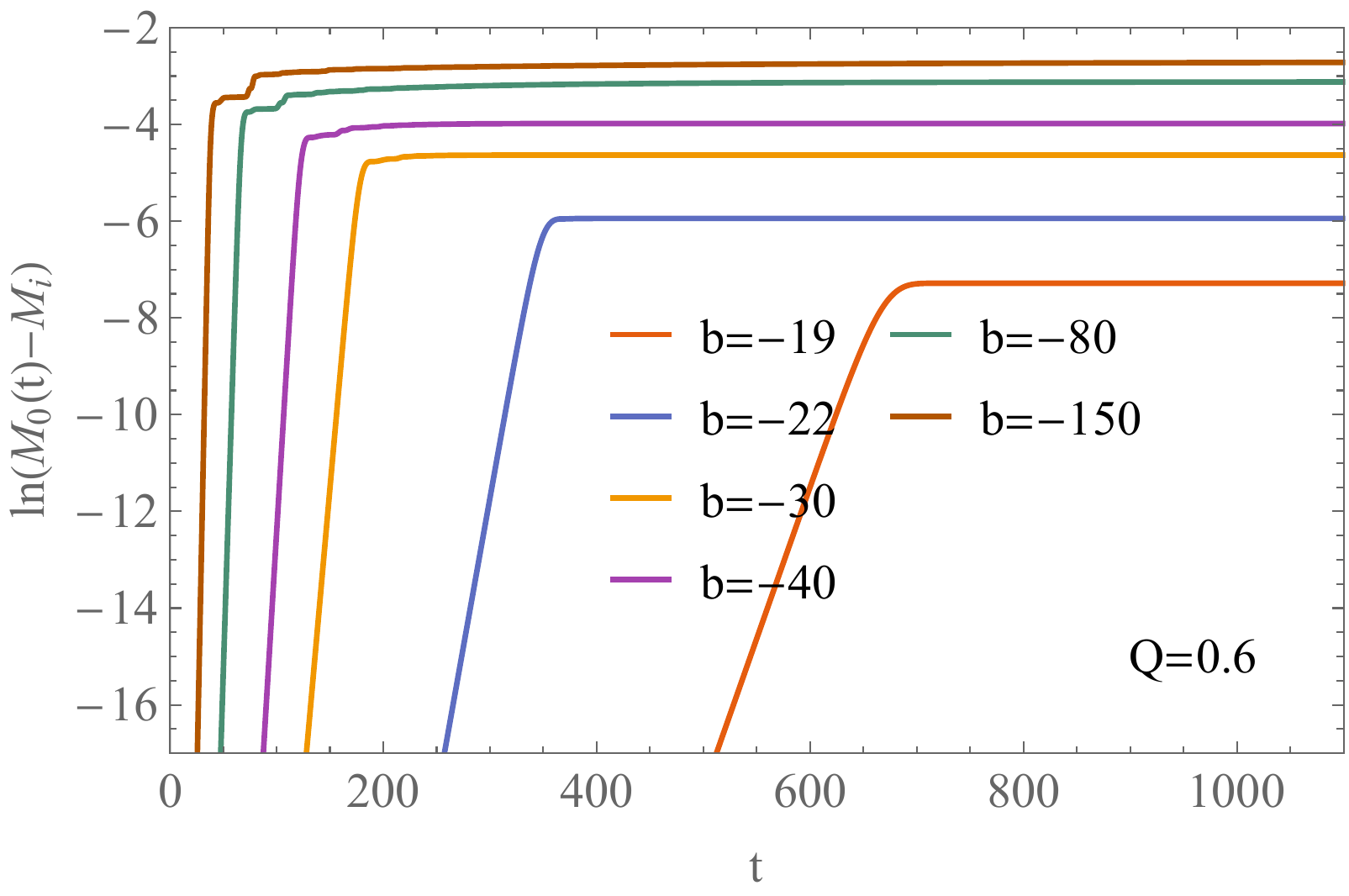}} & {\footnotesize{}\includegraphics[width=0.42\textwidth]{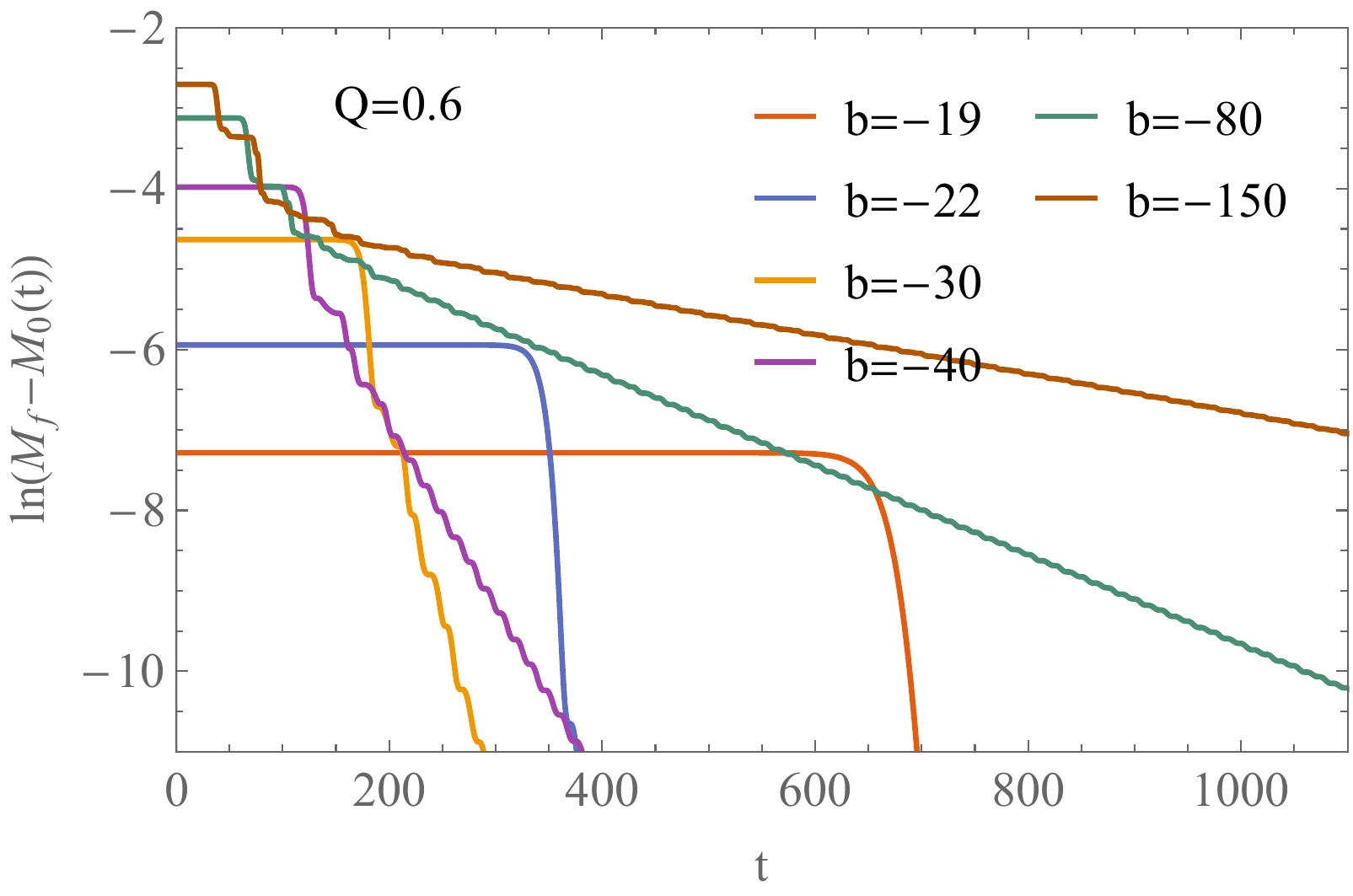}}\tabularnewline
\end{tabular}{\footnotesize\par}
\par\end{centering}
{\footnotesize{}\caption{\label{fig:Q96irrMindex}{\small{} }The evolution of $\ln(M_{0}-M_i)$ (left) and $\ln(M_f-M_{0})$ (right) of the irreducible mass of the black hole for various $b$ when $Q=0.9$ and 0.6. Here $\Lambda=-0.03$.}
}{\footnotesize\par}
\end{figure*}

%But the increment of the irreducible mass during the evolution is larger for black holes with larger charge.
Interestingly, we find that the irreducible mass increases exponentially at early times and
saturates to the final value also exponentially. Namely, the evolution of the irreducible mass has behavior:
\begin{eqnarray}\label{fitting}
M_{0}(t) & \simeq\begin{cases}
M_{i}+e{}^{\gamma_{i}t+c_{i}}, & \text{early times,}\\
M_{f}-e{}^{-\gamma_{f}t+c_{f}}, & \text{late times}.
\end{cases}\label{eq:MirrExp}
\end{eqnarray}
%Fig.\ref{fig:Q96irrMindex} shows evolution of the irreducible mass at early times and late time.
%We fit the curves using the following functions.
Here $\gamma_{i}$ is the growth rate at early times and $\gamma_{f}$  the saturating rate at late times. $c_{i,f}$ are some constant depending on the  parameters. This is shown in Fig.\ref{fig:Q96irrMindex} intuitively.
It is worth mentioning that as zooming in the behavior of $\ln(M_f-M_{0})$ at late times, we find it evolves step-likely and each step takes almost the same time that is independent of $b$.
This is reasonable since the outgoing matter can be bounced back  by the AdS boundary.

{\footnotesize{}}
{\footnotesize\par}

\begin{figure*}
\begin{centering}
{\footnotesize{}}%
\begin{tabular}{cc}
{\footnotesize{}\includegraphics[width=0.42\textwidth]{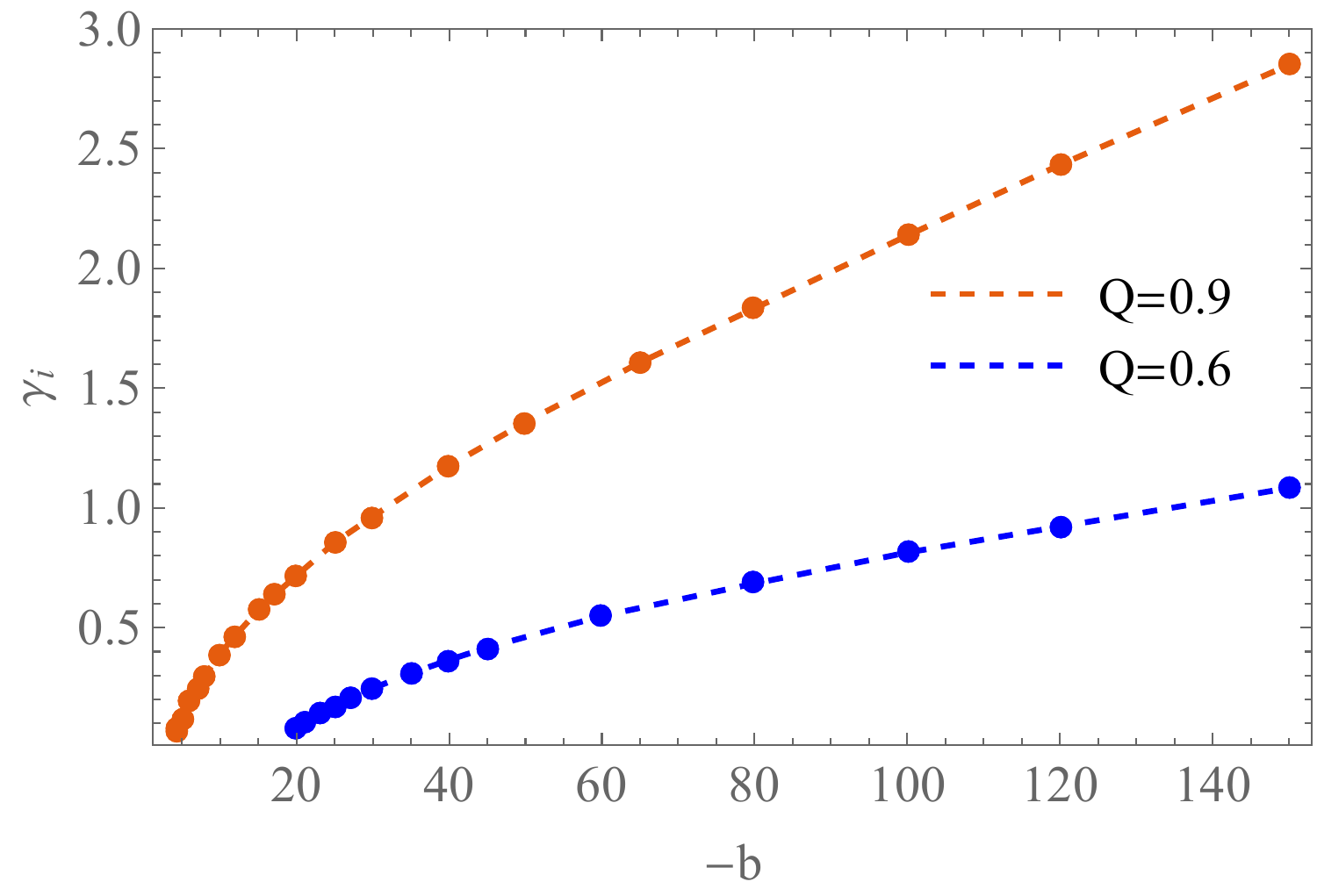}} & {\footnotesize{}\includegraphics[width=0.42\textwidth]{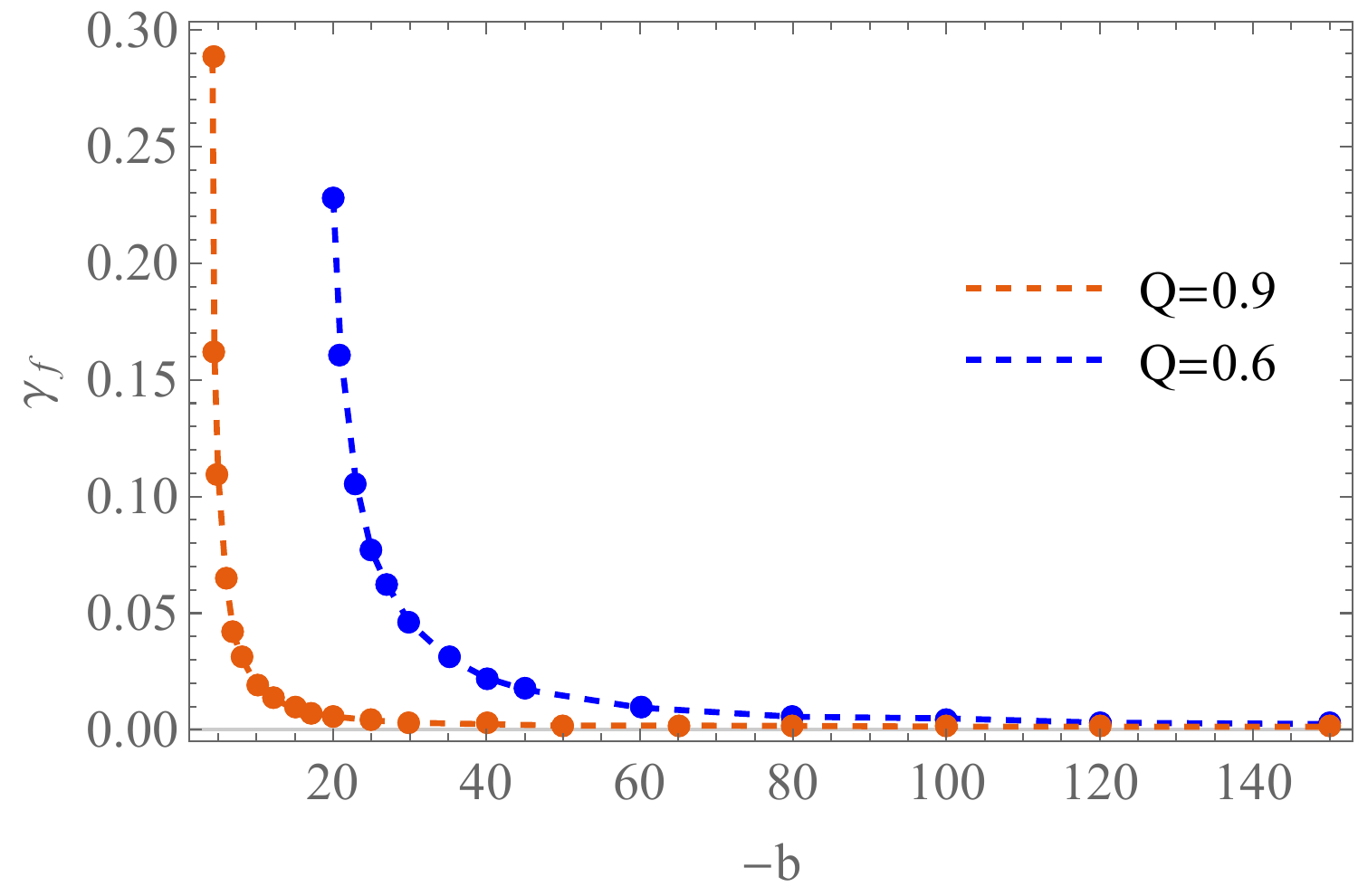}}\tabularnewline
\end{tabular}{\footnotesize\par}
\par\end{centering}
{\footnotesize{}\caption{\label{fig:Q96irrMindexN}{\small{} }The growth rates $\gamma_{i}$
(left) and $\gamma_{f}$ (right) for various $b$ when $Q=0.9$ and
$0.6$. Here $\Lambda=-0.03$.}
}{\footnotesize\par}
\end{figure*}

At early times, the irreducible mass increases faster as $-b$ increases.
At late times, the irreducible mass saturates slower as $-b$ increases.
This means that although the black holes with stronger coupling functions are more likely to be hairy, they take longer time to settle down into hairy black holes. The quantatively dependence of the indexes $\gamma_{i,f}$ on coupling parameter $b$ is shown in Fig.\ref{fig:Q96irrMindexN}.

{\footnotesize{}}
{\footnotesize\par}

\subsubsection{Effects of charge on the black hole spontaneous scalarisation}

\begin{figure*}
\begin{centering}
{\footnotesize{}}%
\begin{tabular}{cc}
 {\footnotesize{}\includegraphics[width=0.45\textwidth]{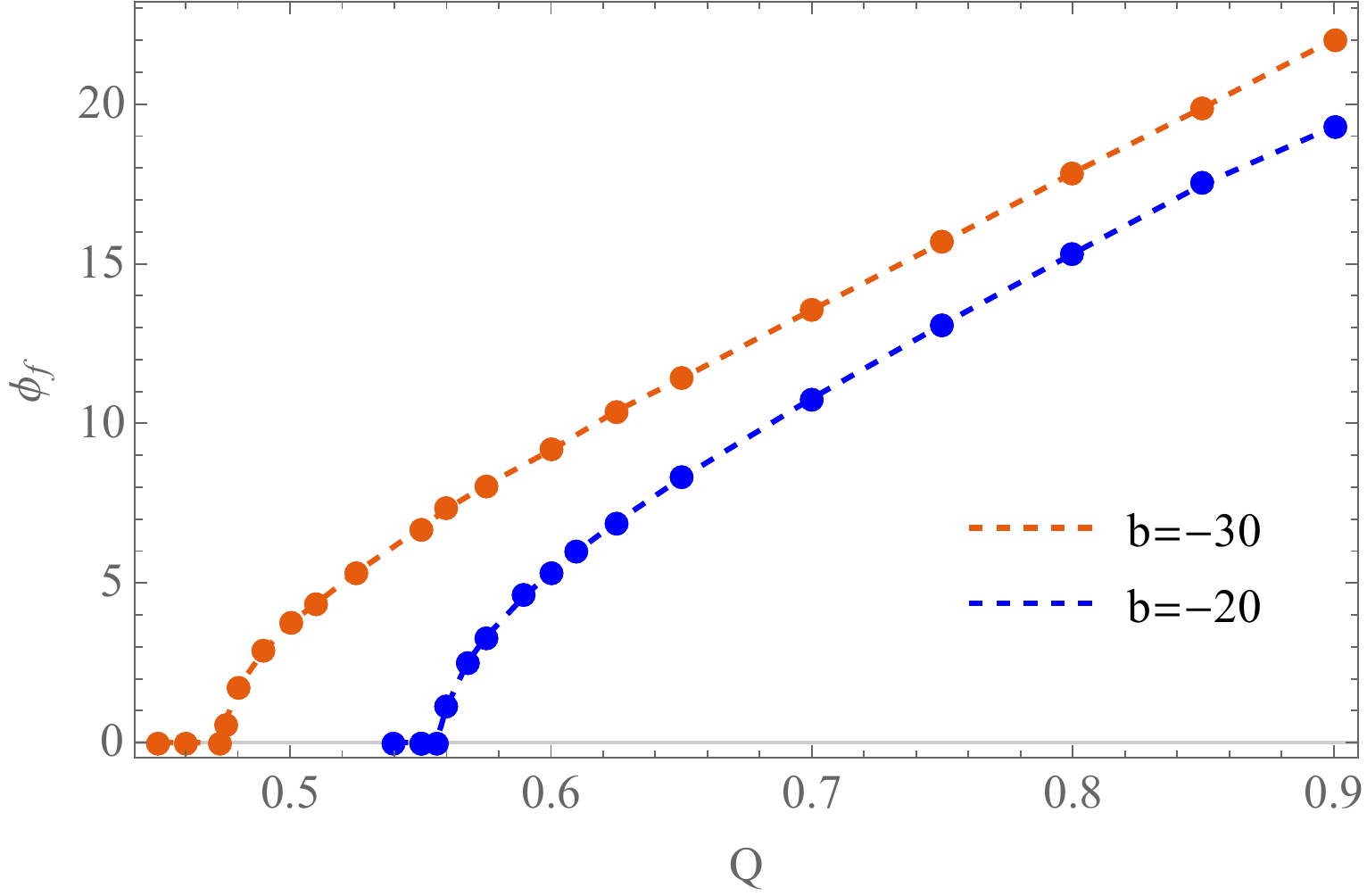}}\tabularnewline
\end{tabular}{\footnotesize\par}
\par\end{centering}
{\footnotesize{}\caption{\label{fig:Q96b3020phi3fr}The final value $\phi_{f}$ of $\phi_{3}(t)$
for various $Q$ when $b=-30$ and $-20$. Here $\Lambda=-0.03$.}
}{\footnotesize\par}
\end{figure*}

In this subsection we  fix $\Lambda=-0.03$, choose $b=-30,-20$ and change $Q$ to study the effects of charge  on the dynamics of the black hole spontaneous scalarisation. The final values  $\phi_{f}$ of $\phi_{3}$ are shown in Fig.\ref{fig:Q96b3020phi3fr}.
There exists a critical charge $Q_{\ast}$ which is about 0.475 for $b=-30$, and $0.55$ for $b=-20$.
The final value $\phi_{f}$  vanishes when $Q<Q_{\ast}$, indicating  a final scalar-free black hole solution.   $\phi_{f}$ has finite value when $Q>Q_{\ast}$, indicating   a scalar hairy black hole solution.
Near the critical value $Q_{\ast}$ where spontaneous scalarisation occurs, $\phi_{f}$ changes unsmoothly.

\begin{figure}
  \centering
  \includegraphics[width=0.6\textwidth]{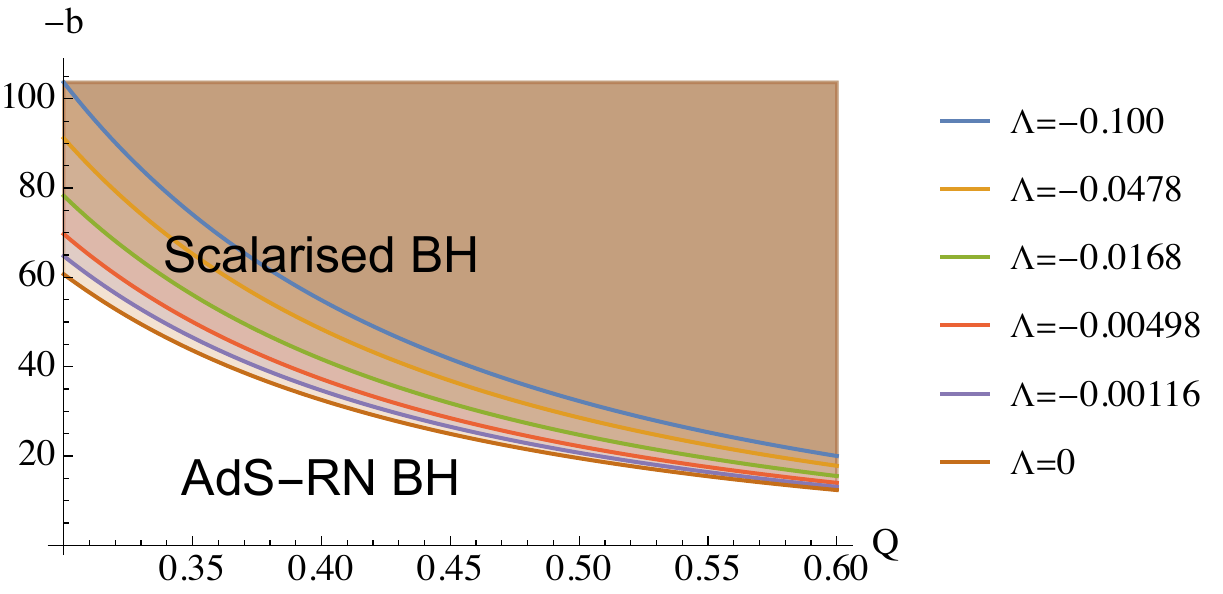}
  \caption{The phase structure along $Q$-direction. The shaded region is the parameter region where scalarisation occurs.}
  \label{fig:phasealongq}
\end{figure}

	We present the phase structure of the scalarisation along $Q$-direction in Fig. \ref{fig:phasealongq}. Apparently, the critical $-b$ decreases when $Q$ increases, suggesting that the system can scalarise more easily with larger $Q$.

{\footnotesize{}}
{\footnotesize\par}

\begin{figure*}
\begin{centering}
{\footnotesize{}}%
\begin{tabular}{cc}
{\footnotesize{}\includegraphics[width=0.42\textwidth]{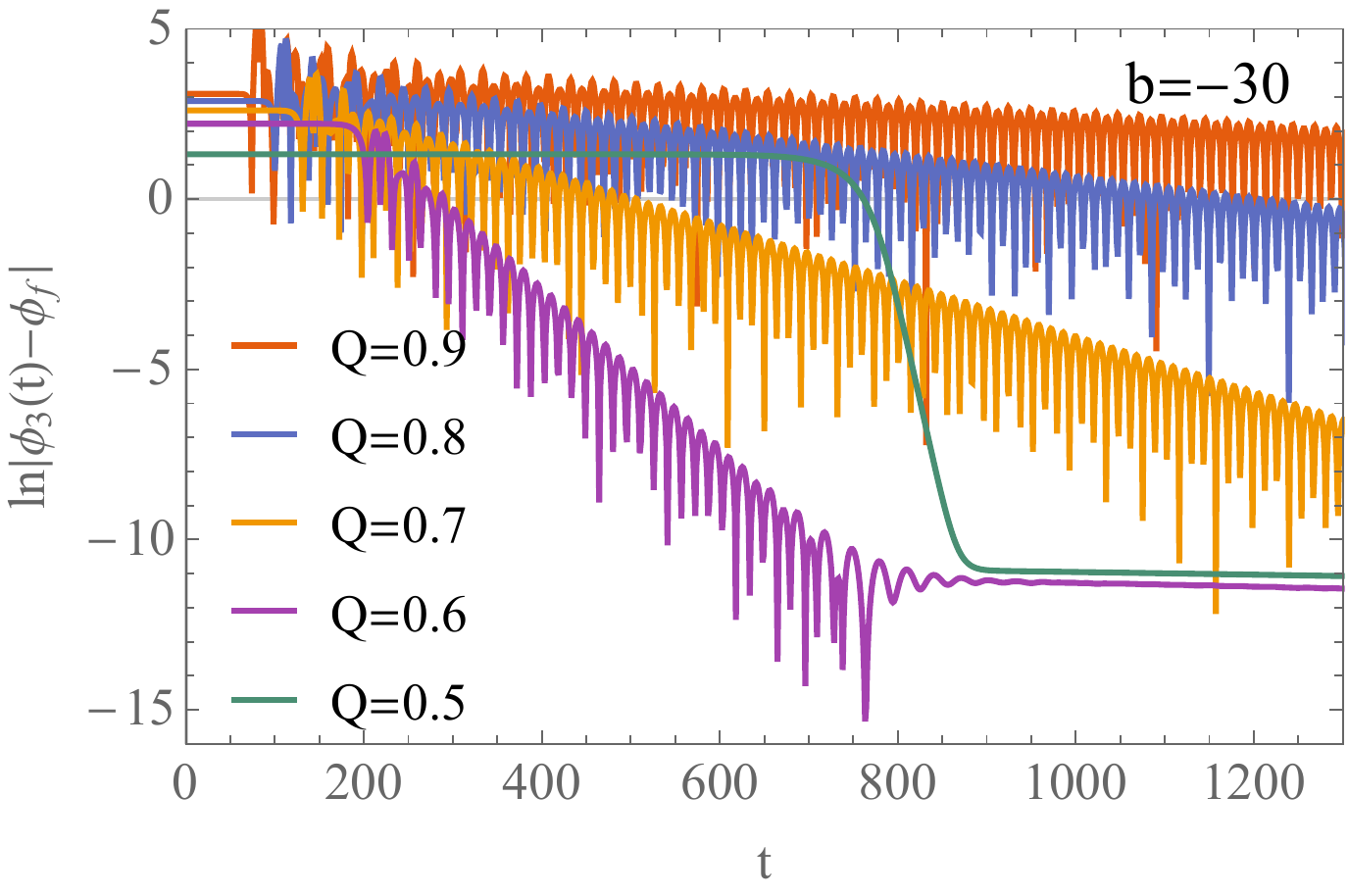}} & {\footnotesize{}\includegraphics[width=0.42\textwidth]{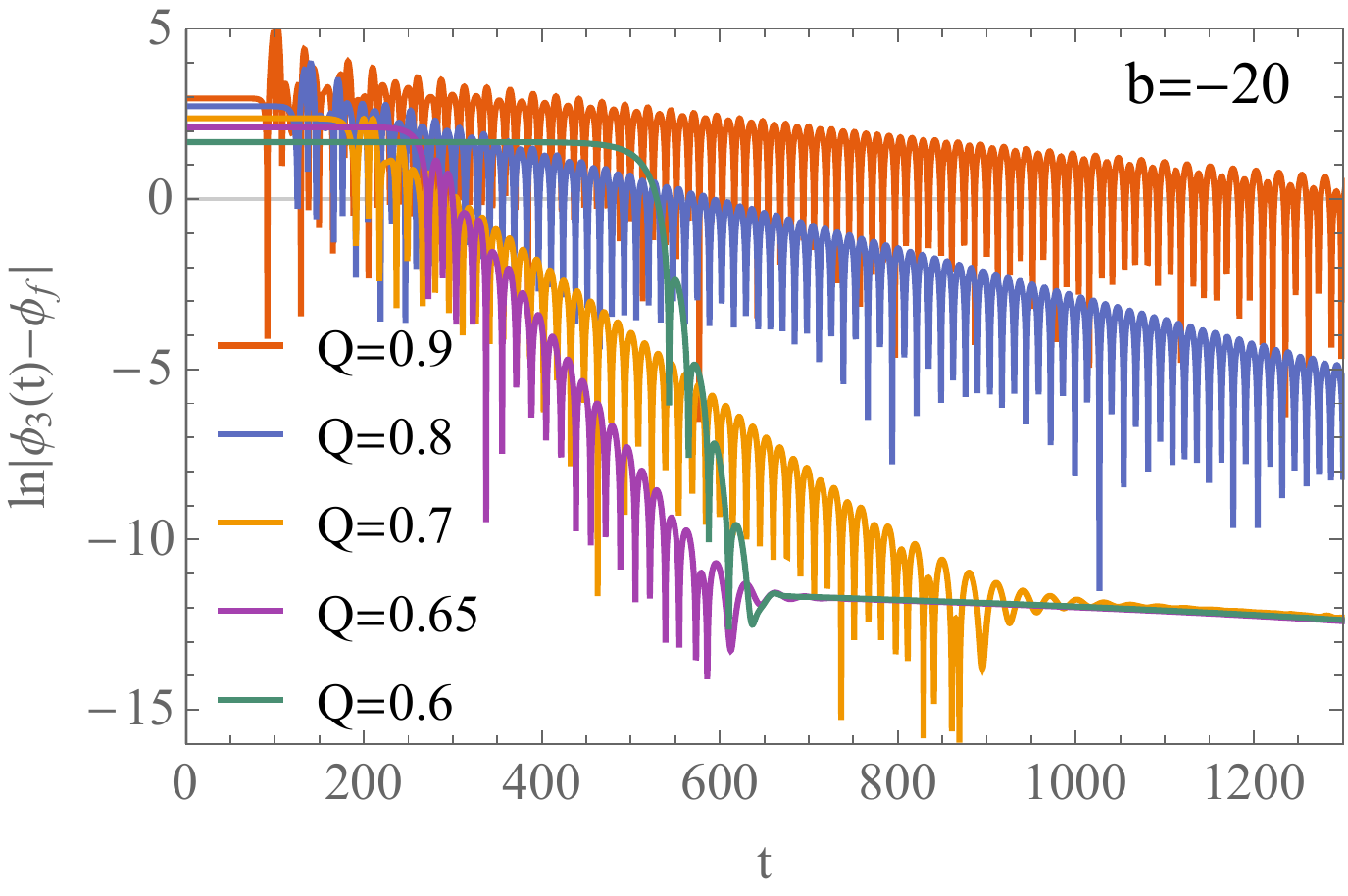}}\tabularnewline
{\footnotesize{}\includegraphics[width=0.42\textwidth]{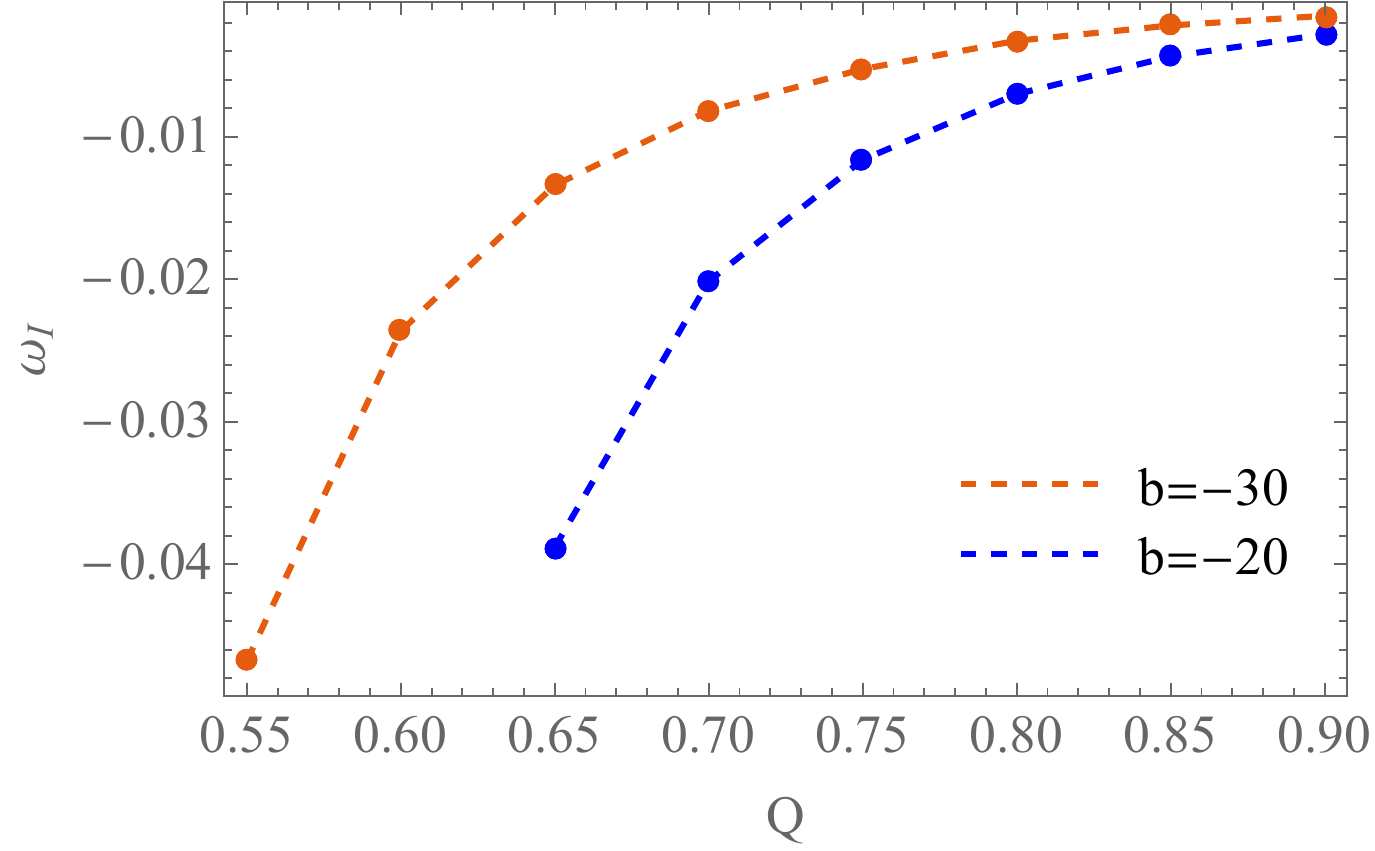}} & {\footnotesize{}\includegraphics[width=0.42\textwidth]{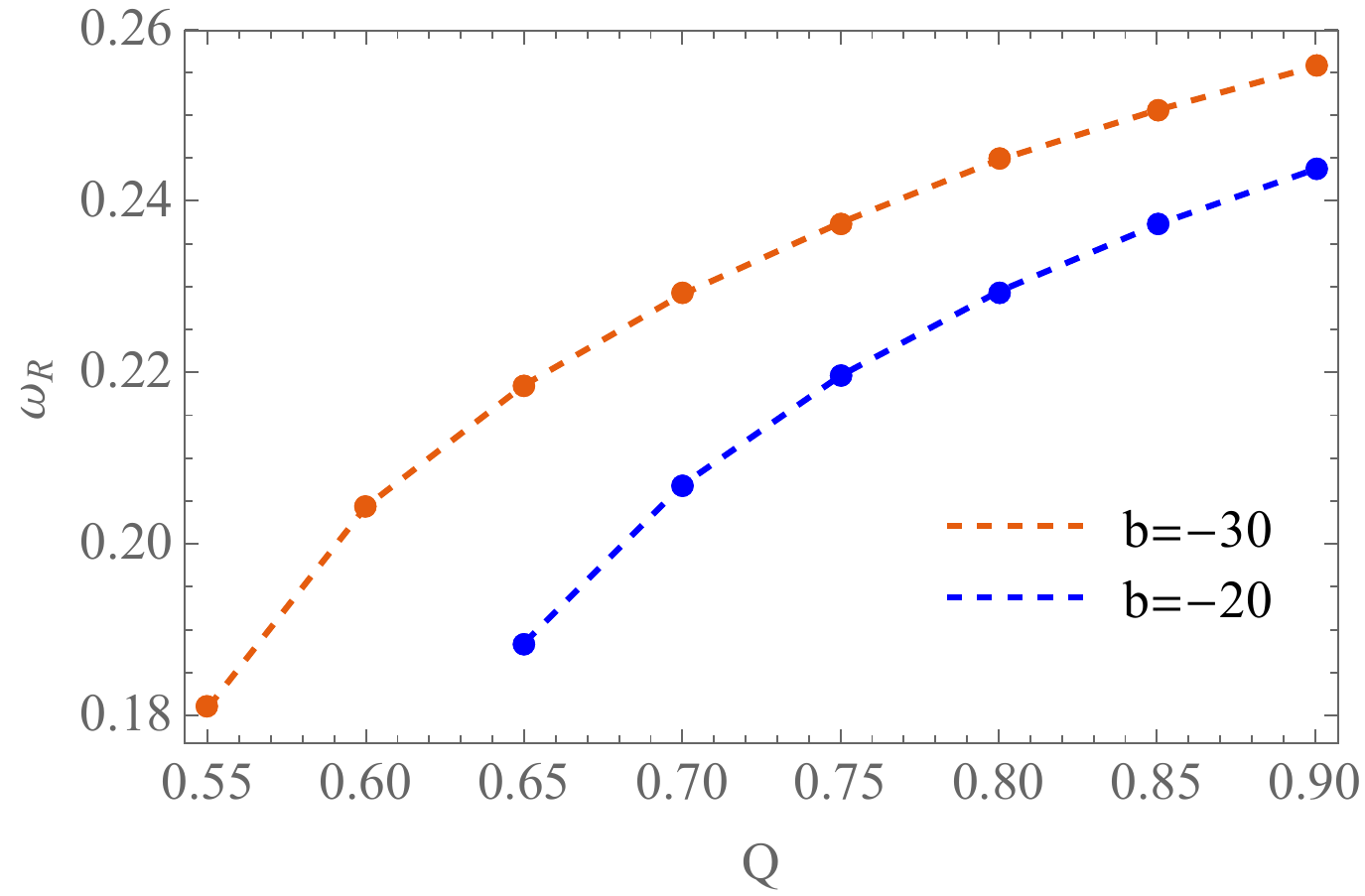}}\tabularnewline
\end{tabular}{\footnotesize\par}
\par\end{centering}
{\footnotesize{}\caption{\label{fig:b3020phi3wIR}{\small{} }The evolution of $\ln|\phi_{3}-\phi_f|$ (upper) and the complex frequencies dominant damping modes of $\phi_{3}$ v.s. $Q$ when $b=-30$
and $-20$. Here $\Lambda=-0.03$.}
}{\footnotesize\par}
\end{figure*}

The evolution of $\phi_{3}$ with $b=-30$ and $-20$ is shown in the upper panels of Fig. \ref{fig:b3020phi3wIR}. Using  discrete Fourier transformation or Prony method, we can work out the complex frequencies of the component modes.
The  zero mode with $\omega_R=0$ corresponding to nonvanishing $\phi_f$ exists only when $Q>Q_{\ast}$.
The real and imaginary parts of the complex frequencies of the dominant damping modes are plotted in the lower panels of Fig.\ref{fig:b3020phi3wIR}.
As $Q$ increases, $\omega_{I}$ tends to zero and $\omega_R$ increases monotonically.
This implies the system with larger charge oscillates faster and needs more time to settle down to be a static hairy black hole.

{\footnotesize{}}
{\footnotesize\par}

\begin{figure*}
\begin{centering}
{\footnotesize{}}%
\begin{tabular}{cc}
{\footnotesize{}\includegraphics[width=0.42\textwidth]{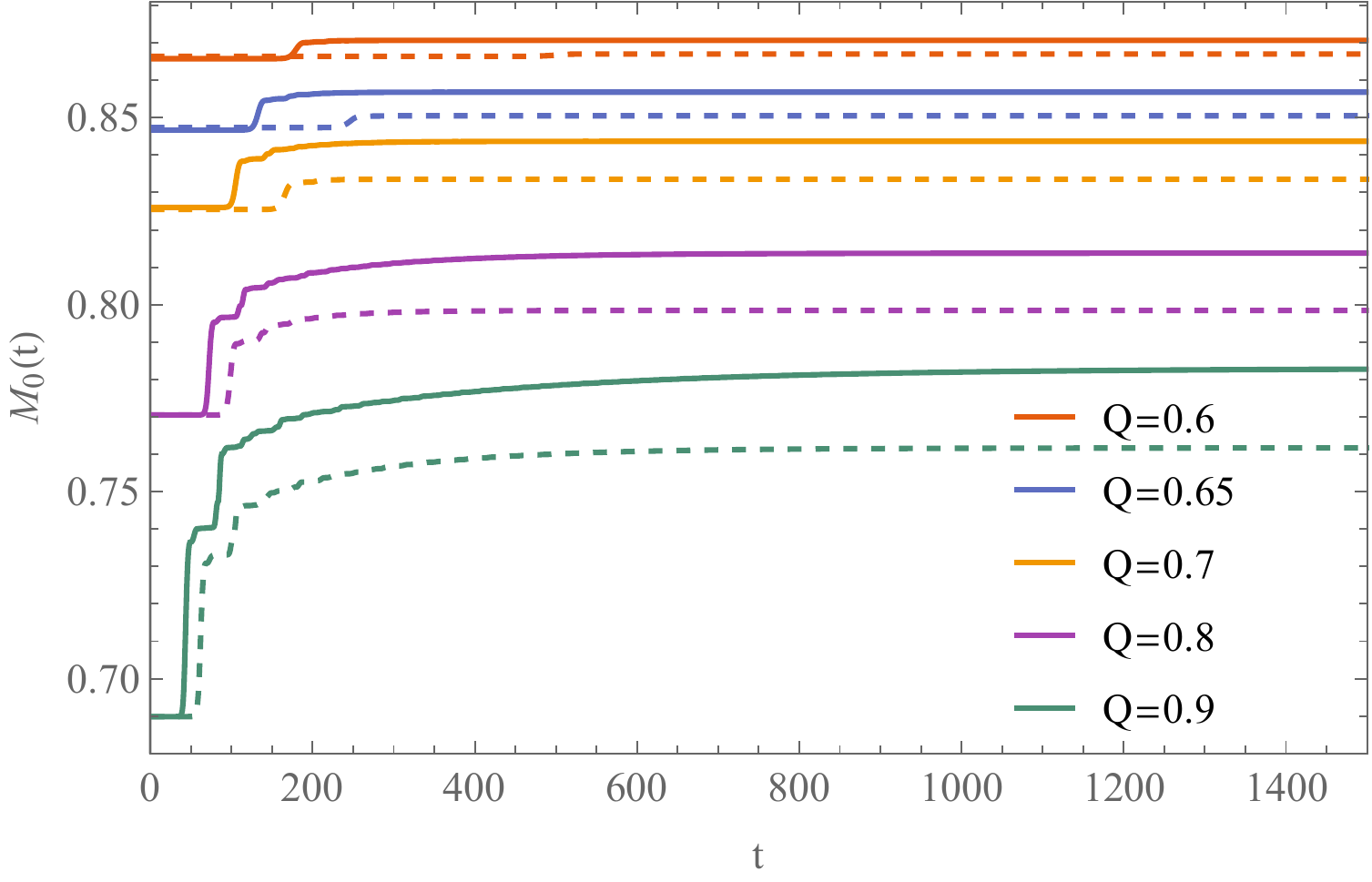}} & {\footnotesize{}\includegraphics[width=0.42\textwidth]{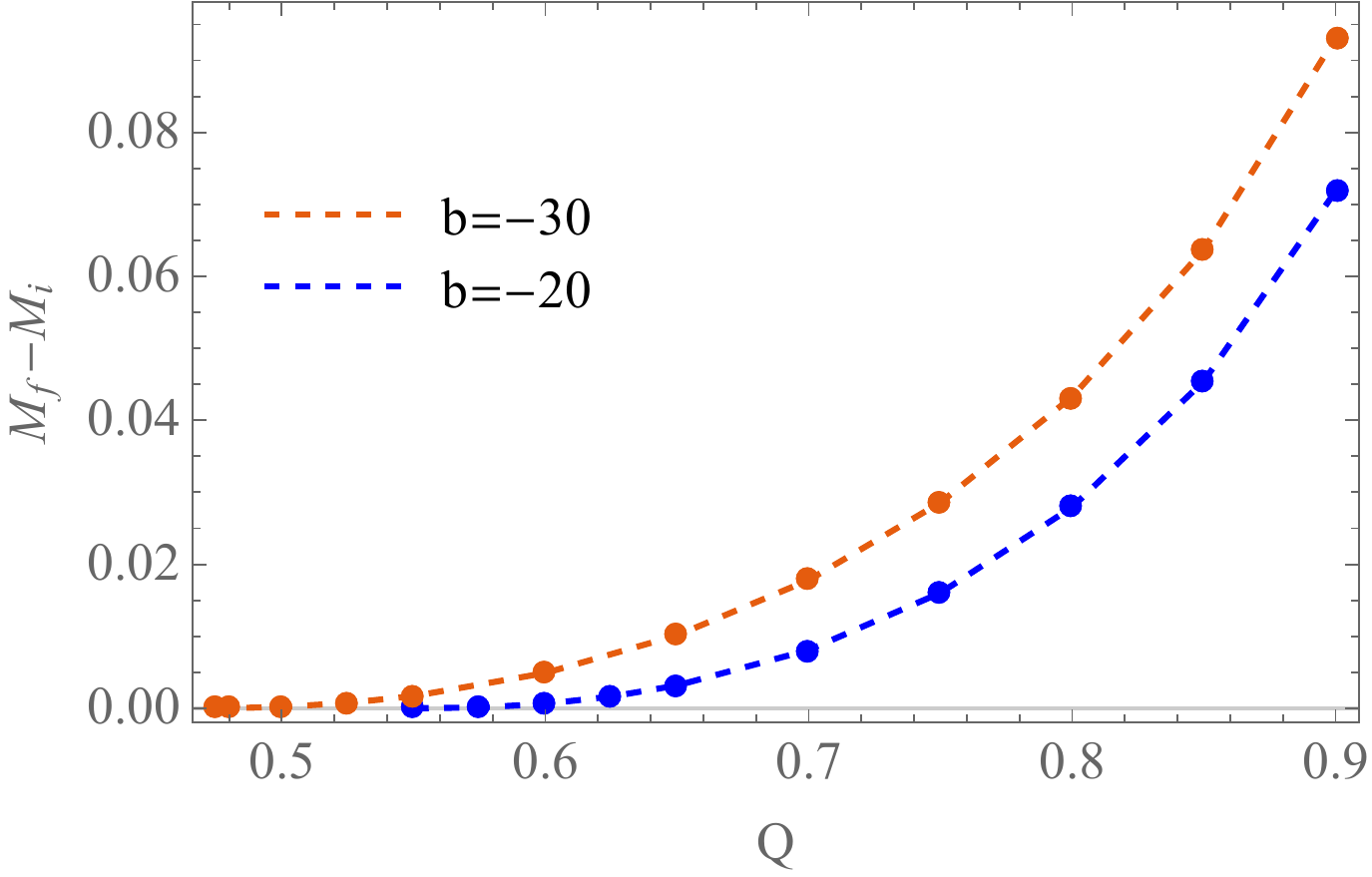}}\tabularnewline
\end{tabular}{\footnotesize\par}
\par\end{centering}
{\footnotesize{}\caption{\label{fig:b3020irrM}{\small{} }Left panel show the evolution of
irreducible mass of the black hole for various $Q$ when $b=-30$
(solid lines) and $-20$ (dashed lines). Right panel for the final
value of the irreducible mass v.s. $Q$.}
}{\footnotesize\par}
\end{figure*}

The evolution of the irreducible mass $M_{0}$ is shown in the left of Fig.\ref{fig:b3020irrM}.
Note that $M_{0}$ never decreases in the evolution.
As $Q$ increases, both the initial and final values of $M_{0}$ decreases.
We plot the increment of the irreducible mass $M_{f}-M_{i}$ in the right of Fig. \ref{fig:b3020irrM}.
The irreducible mass of the black hole with larger charge increases more.

{\footnotesize{}}
{\footnotesize\par}

\begin{figure*}
\begin{centering}
{\footnotesize{}}%
\begin{tabular}{cc}
{\footnotesize{}\includegraphics[width=0.42\textwidth]{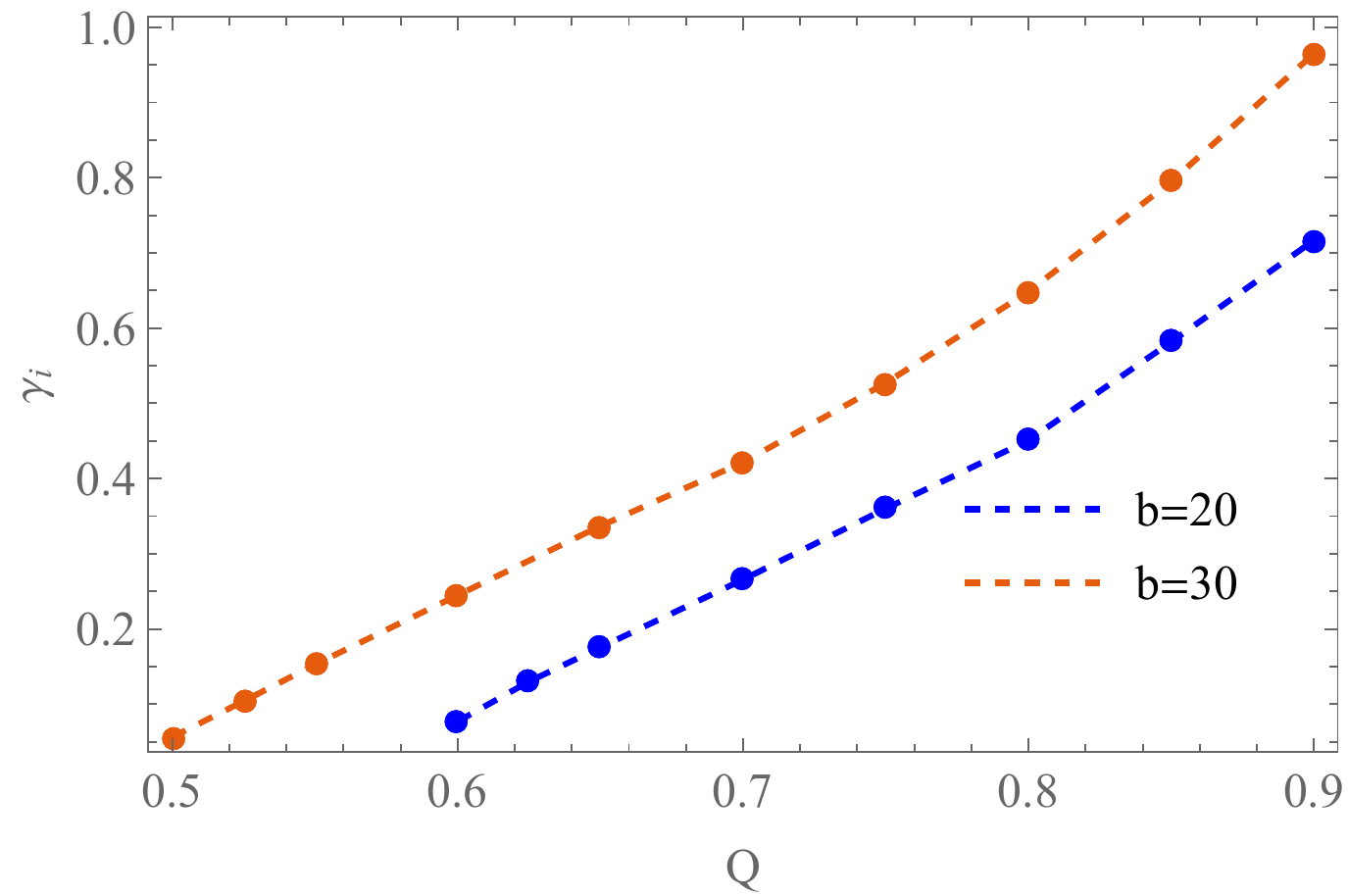}} & {\footnotesize{}\includegraphics[width=0.42\textwidth]{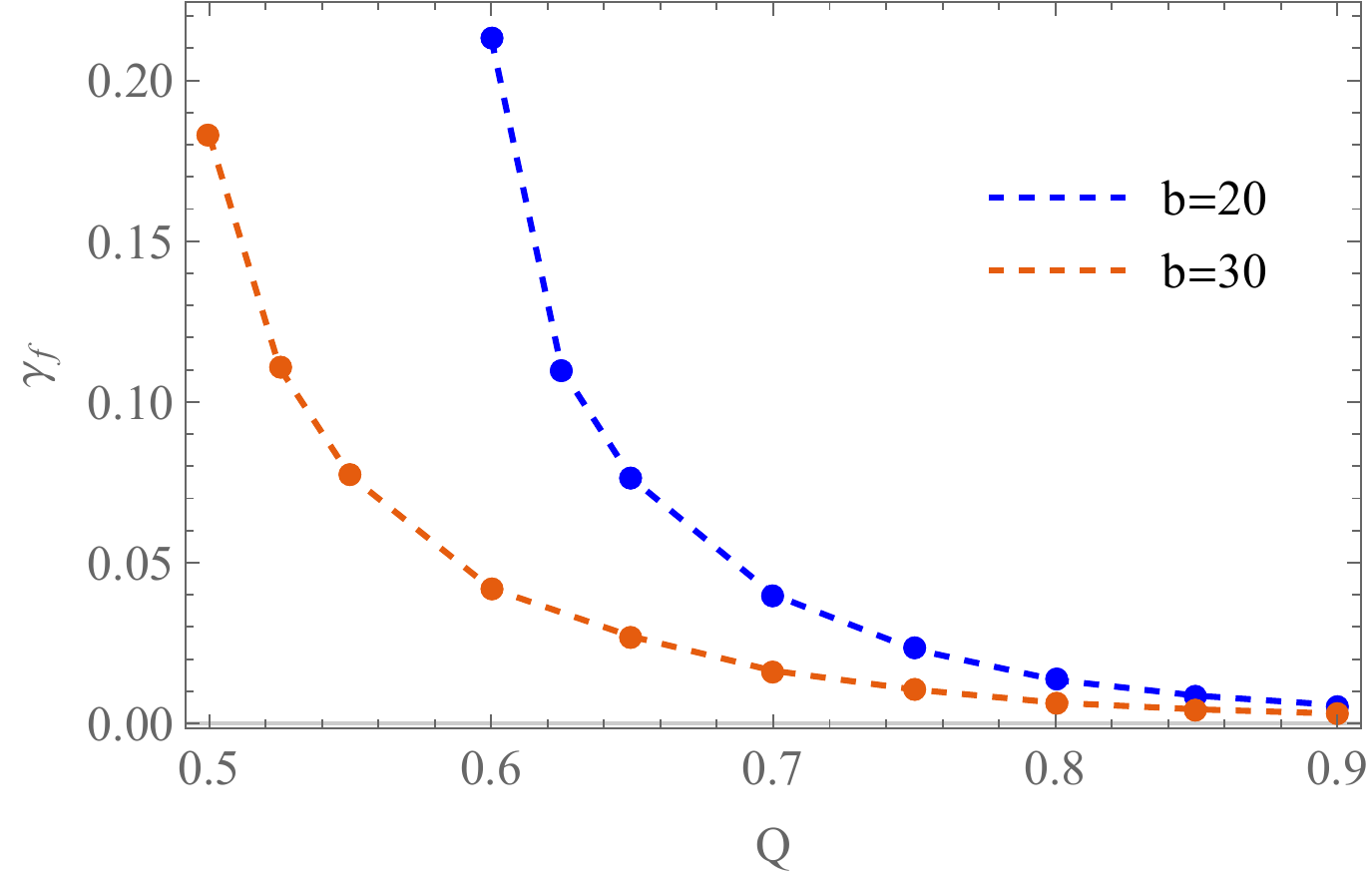}}\tabularnewline
\end{tabular}{\footnotesize\par}
\par\end{centering}
{\footnotesize{}\caption{\label{fig:b3020irrMindexN}{\small{} }The growth rates $\gamma_{i}$
(left) and $\gamma_{f}$ (right) for various $Q$ when $b=-30$ and
$-20$.}
}{\footnotesize\par}
\end{figure*}

The evolution of the irreducible mass still behaves as (\ref{eq:MirrExp}) at early times and late times.
%We data fit the irreducible mass with (\ref{eq:MirrExp}) at early times and late times.
The indexes $\gamma_{i,f}$ are shown in Fig.\ref{fig:b3020irrMindexN}.
At early times, the index $\gamma_{i}$ is nearly proportional to the charge and the slope is almost independent of $b$.
The irreducible mass increases faster as $Q$ increases.
At late times, the index $\gamma_{f}$ tends to zero as charge increases.
The irreducible mass saturates slower to the final value. This means that the black holes with larger charge are slower to settle down to be the final static hairy black holes.

{\footnotesize{}}
{\footnotesize\par}

\subsection{Effects of the cosmological constant on  spontaneous scalarisation}

\begin{figure*}
\begin{centering}
{\footnotesize{}}%
\begin{tabular}{cc}
{\footnotesize{}\includegraphics[width=0.42\textwidth]{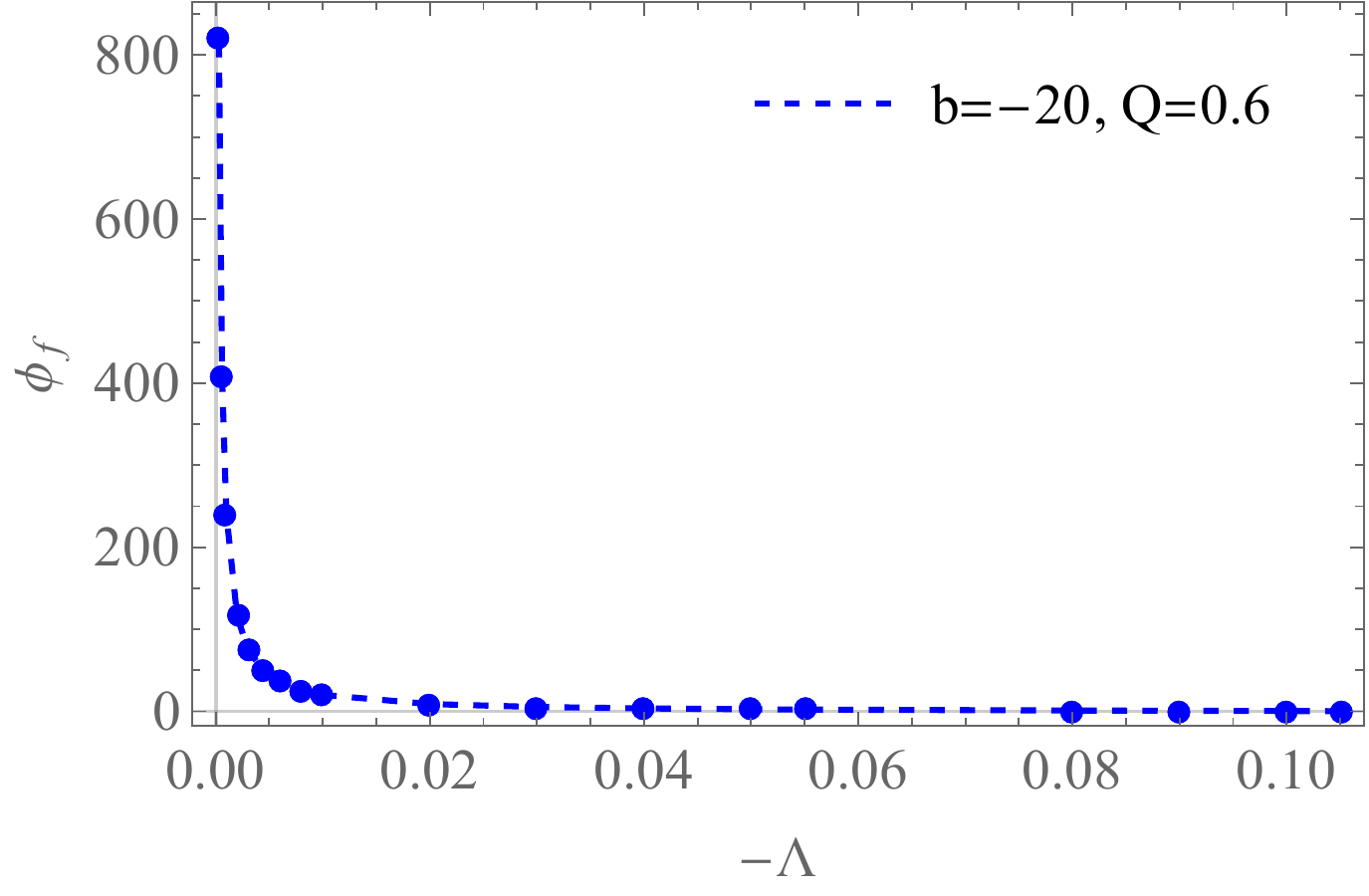}} & {\footnotesize{}\includegraphics[width=0.42\textwidth]{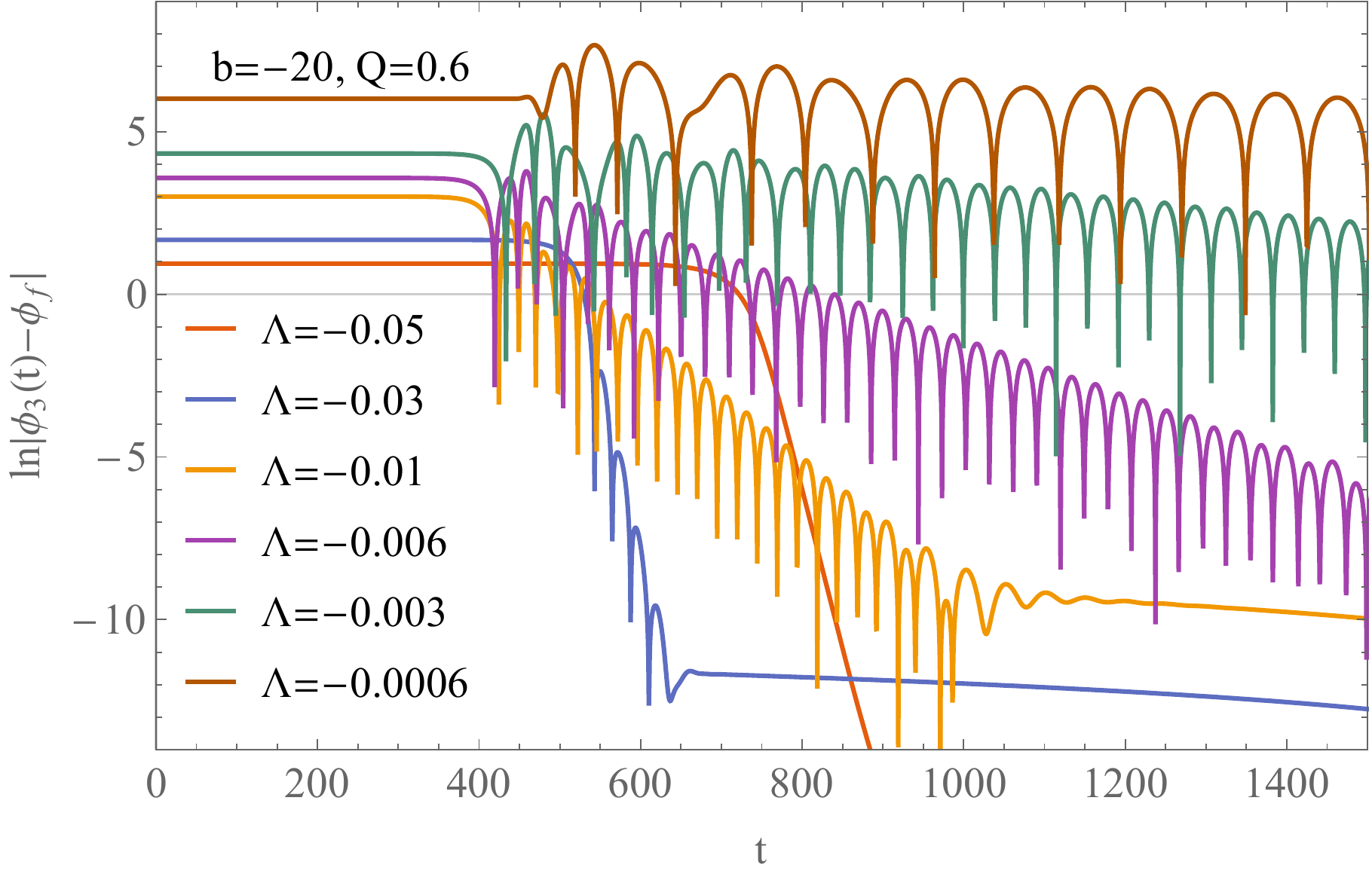}}\tabularnewline
{\footnotesize{}\includegraphics[width=0.42\textwidth]{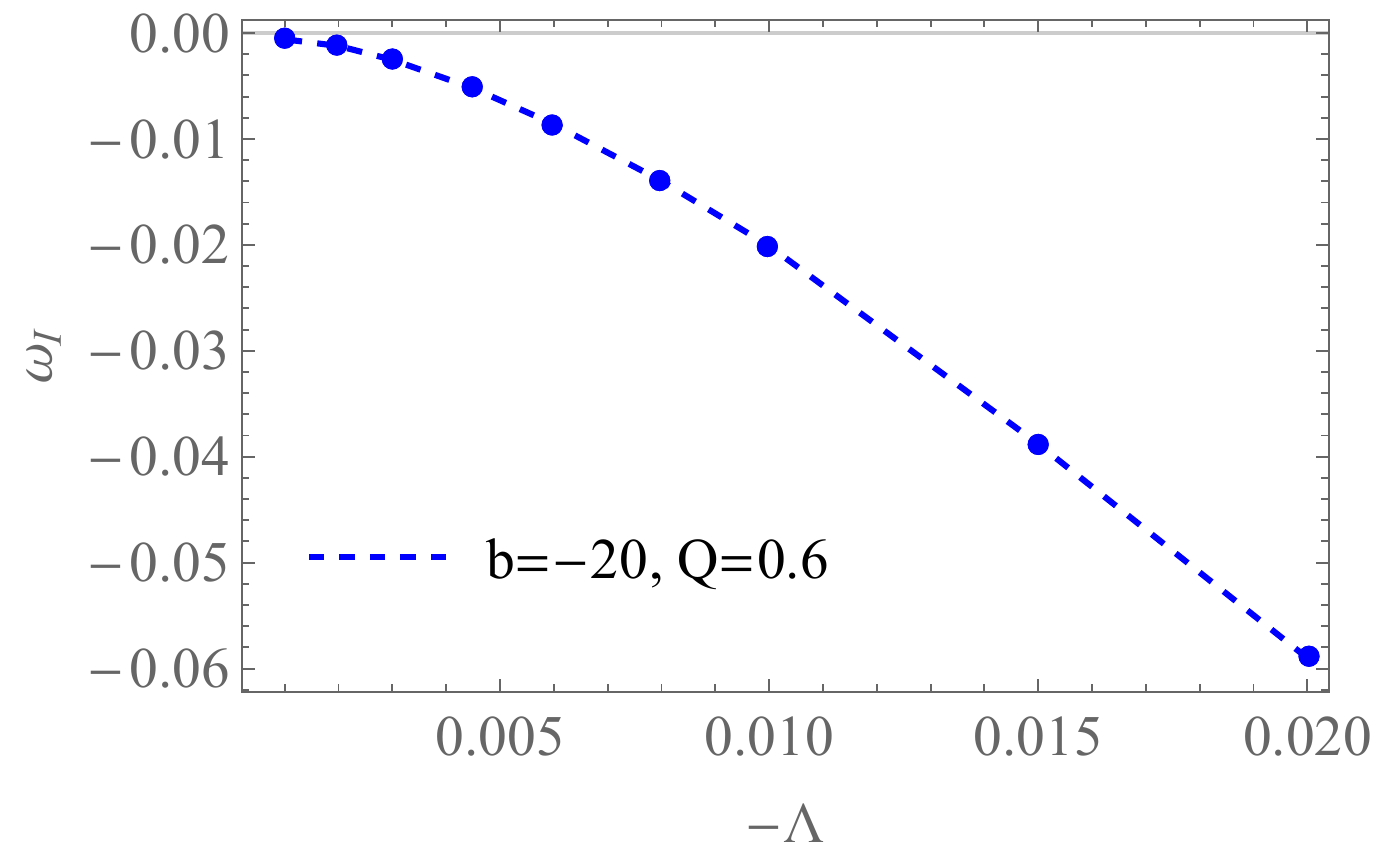}} & {\footnotesize{}\includegraphics[width=0.42\textwidth]{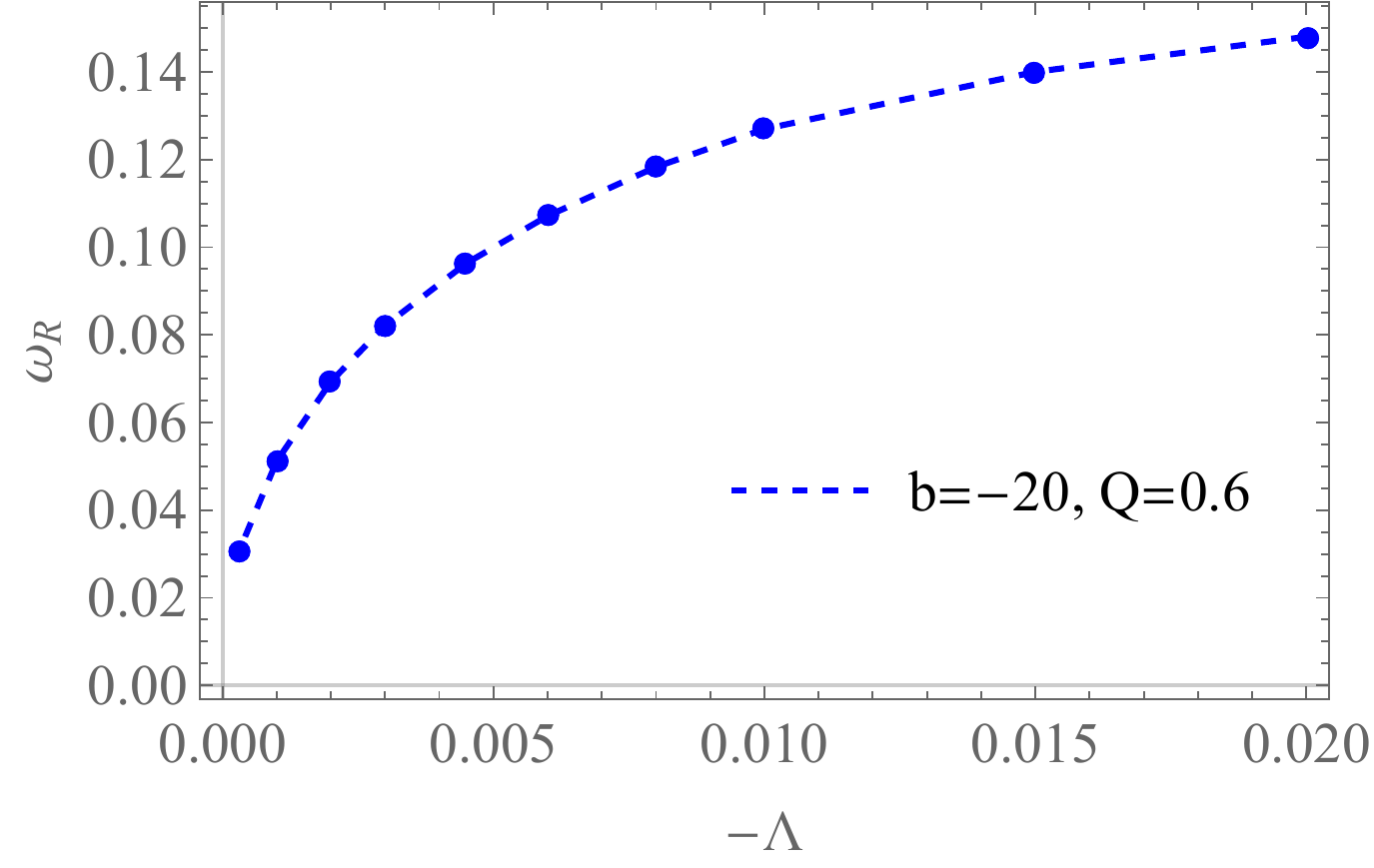}}\tabularnewline
\end{tabular}{\footnotesize\par}
\par\end{centering}
{\footnotesize{}\caption{\label{fig:LPhi3}{\small{} }The final values of $\phi_{f}$ of $\phi_{3}$
(upper left), the evolution of $\ln|\phi_{3}-\phi_{f}|$ (upper right)
and the complex frequencies of the dominant damping modes of $\phi_{3}$
(lower panels) for various $\Lambda$.}
}{\footnotesize\par}
\end{figure*}

In this subsection, we fix $Q=0.6$ and $b=-20$ and change $\Lambda$
to study the effects of the cosmological constant on the black hole
evolution. We show the final value $\phi_{f}$ of $\phi_{3}$ in the
upper left panel of Fig.\ref{fig:LPhi3}. The final static hairy
black hole solution with $\phi_{f}>0$ exists only when $\Lambda$
is larger than a critical value $\Lambda_{\ast}\simeq-0.105$. The
final value $\phi_{f}$ is proportional to $-\Lambda^{-1}$ as $\Lambda\to0$.
This implies that the asymptotic boundary solution in AdS spacetime
can not be generalized straightforwardly to the asymptotic flat spacetime.
In fact, the asympotitcal expansion of the scalar behaves as $\phi\sim c+\phi_{1}/r+\mathcal{O}(r^{-2})$
in asymptotical flat spacetime. In the upper right panel of Fig.\ref{fig:LPhi3}
we shown the evolution of $\ln|\phi_{3}-\phi_{f}|$ for various $\Lambda$.
It osicillates faster and damps faster as $-\Lambda$ increases. The
complex frequencies of the dominant damping modes of $\phi_{3}$ are
shown in the lower panels of Fig.\ref{fig:LPhi3}. As $-\Lambda$
increases, $\omega_{I}$ decreases and $\omega_{R}$ increases. Both
the imaginary part of the frequency $\omega_{I}$ and the real part
$\omega_{R}$ tend to zero as $\Lambda\to0$. The system needs much
more time to settle down for small $-\Lambda$.

\begin{figure}
  \centering
  \includegraphics[width=0.6\textwidth]{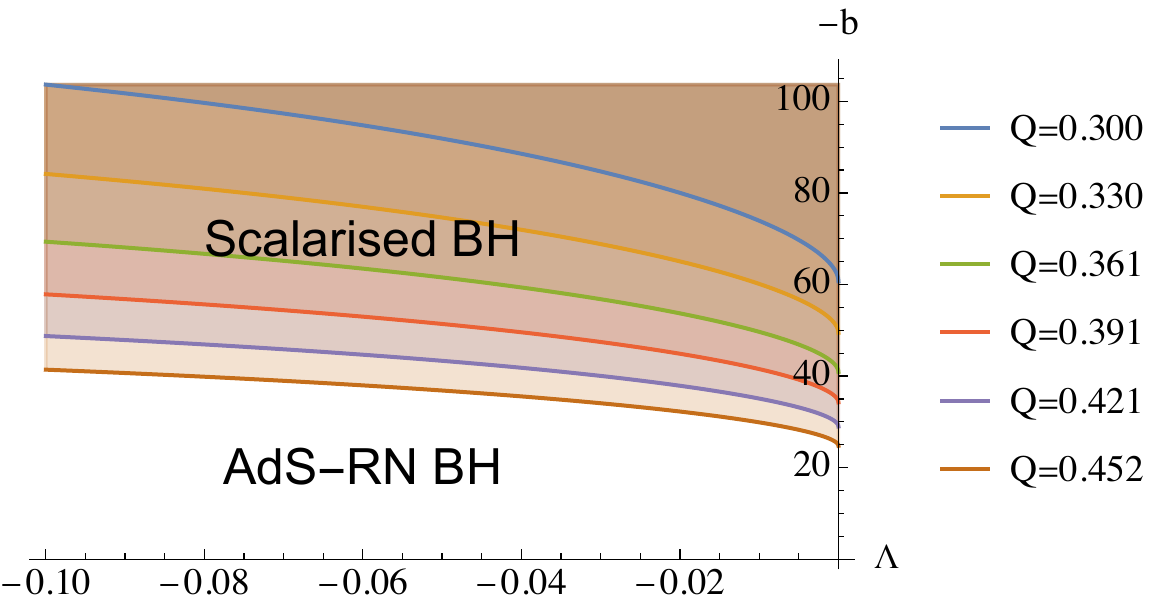}
  \caption{The phase structure along $\Lambda$-direction. The shaded region is the parameter region where scalarisation occurs.}
  \label{fig:phasealonglambda}
\end{figure}

	We present the phase structure of the scalarisation along $\Lambda$-direction in Fig. \ref{fig:phasealonglambda}. Apparently, down to $\Lambda = 0$, the critical $-b$ decreases when $\Lambda$ increases, suggesting that the system can scalarise more easily when the constant curvature becomes smaller. Also, the flat case $\Lambda =0$ scalarises more easily than the AdS case $\Lambda <0$, this is in contrast to the result in \cite{Guo:2021zed}.

{\footnotesize{}}
{\footnotesize\par}

\begin{figure*}
\begin{centering}
{\footnotesize{}}%
\begin{tabular}{cc}
{\footnotesize{}\includegraphics[width=0.42\textwidth]{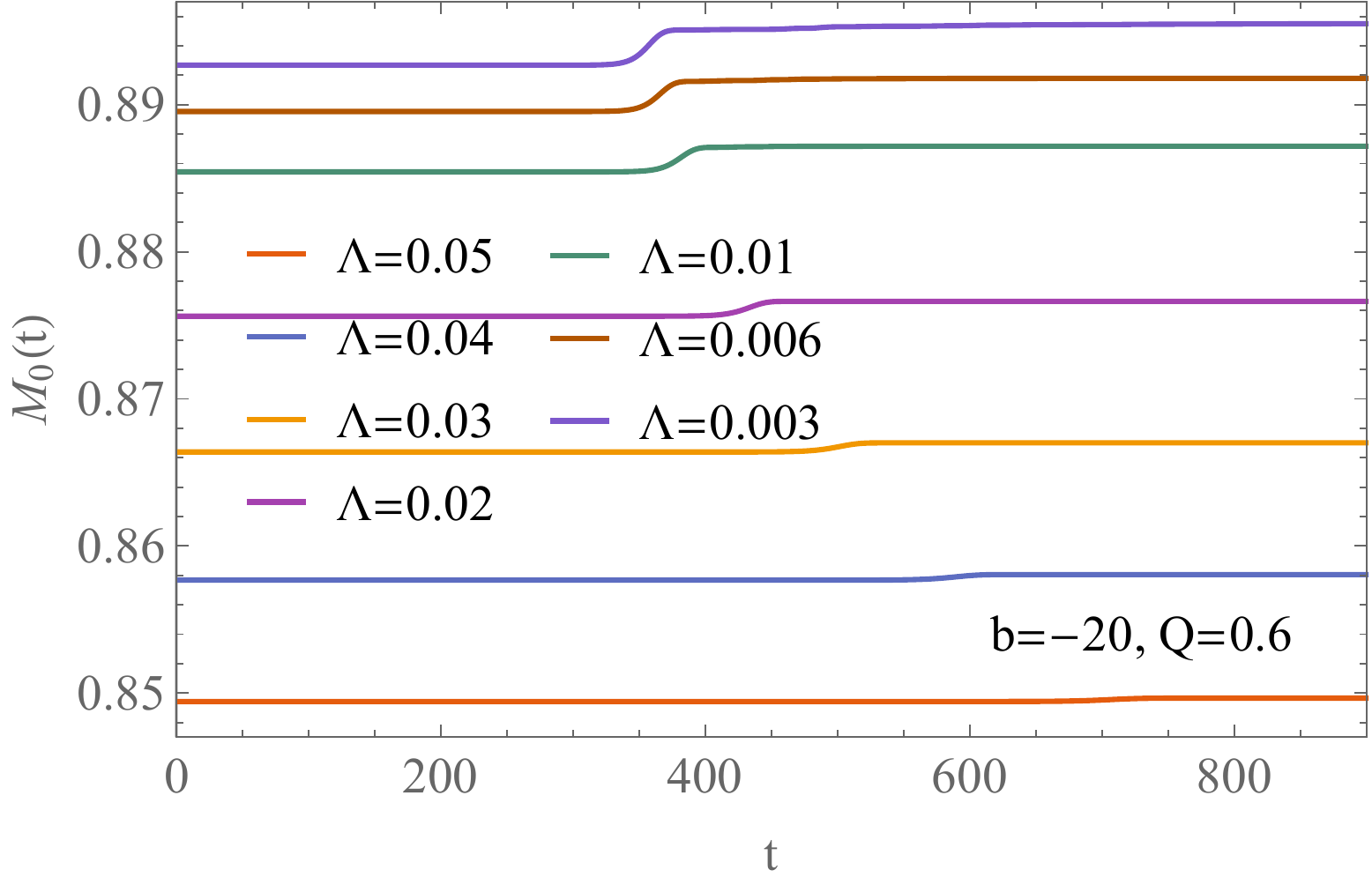}} & {\footnotesize{}\includegraphics[width=0.42\textwidth]{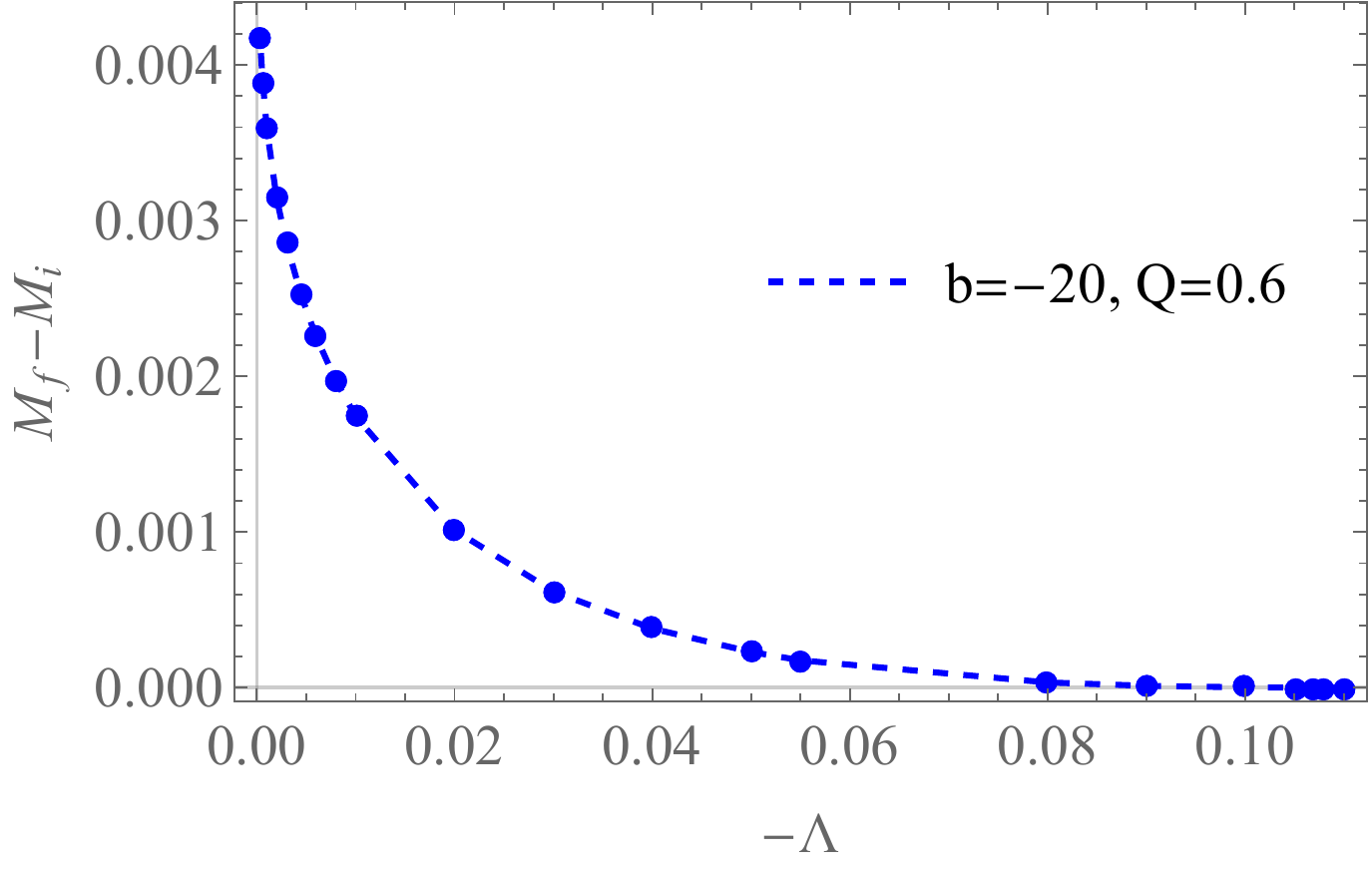}}\tabularnewline
{\footnotesize{}\includegraphics[width=0.42\textwidth]{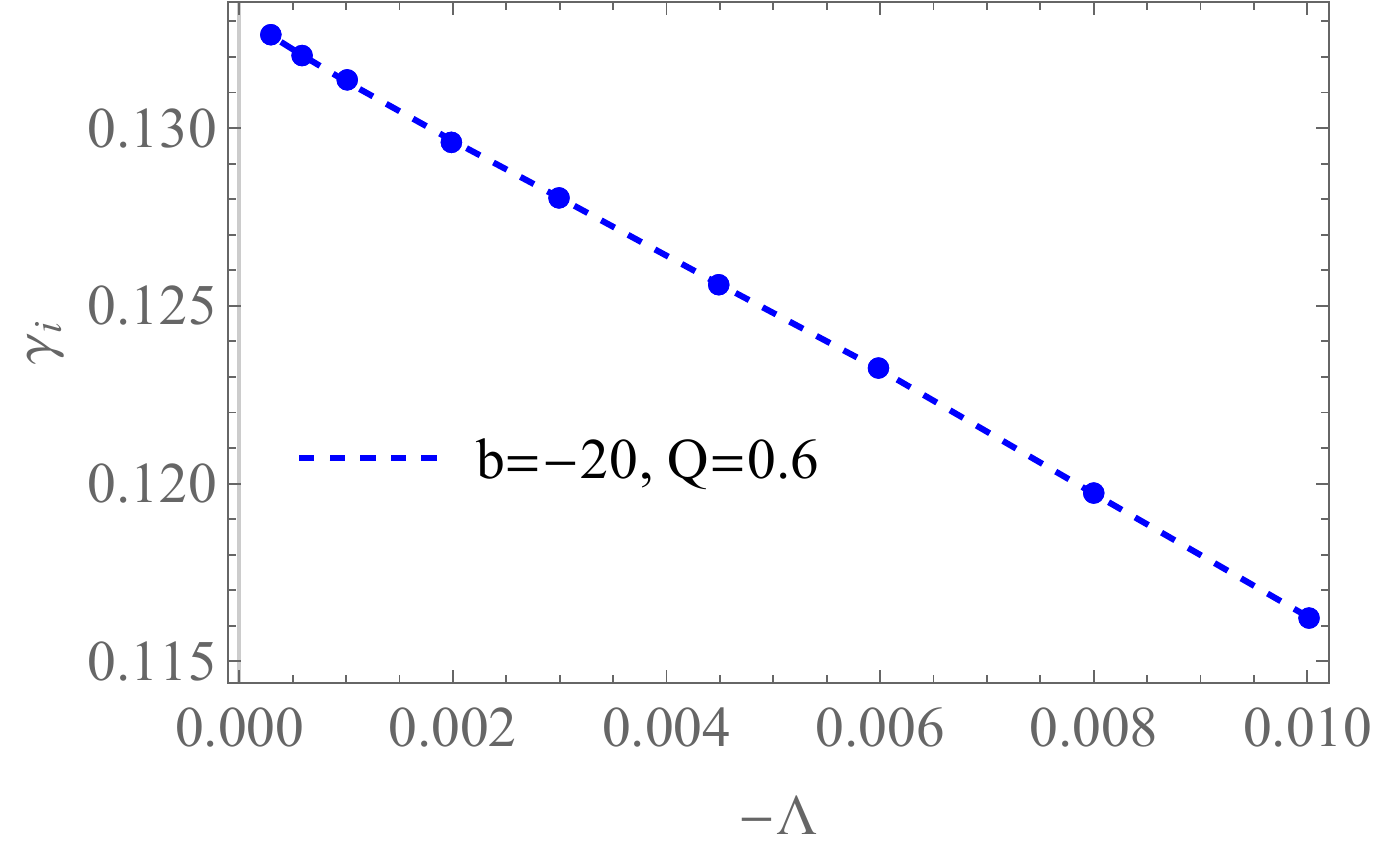}} & {\footnotesize{}\includegraphics[width=0.42\textwidth]{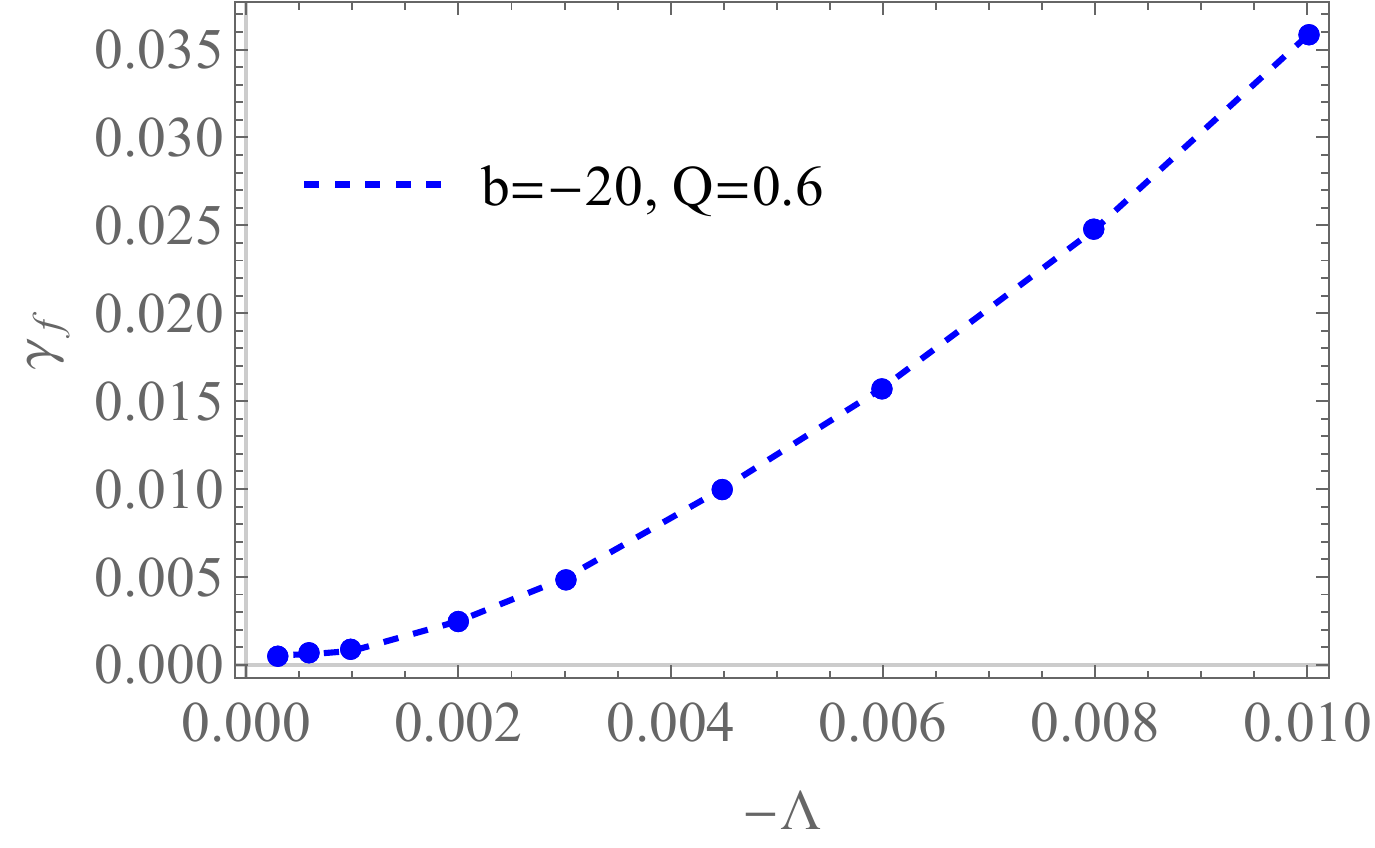}}\tabularnewline
\end{tabular}{\footnotesize\par}
\par\end{centering}
{\footnotesize{}\caption{\label{fig:LM}{\small{} }The evolution of the irreduciable mass $M_{0}$
(upper left), the increasement of the irreduciable mass $M_{f}-M_{i}$
(upper right), the growth rate $\gamma_{i}$ at early times (lower
left) and the saturation rate $\gamma_{f}$ at late times (lower right)
for various $\Lambda$ when $Q=0.6$ and $b=-20$.}
}{\footnotesize\par}
\end{figure*}

The evolution and increasement of the irreduciable mass of the black
hole are shown in the upper panels of Fig.\ref{fig:LM}. The increasement
$M_{f}-M_{i}$ becomes smaller as $-\Lambda$ increases. When $\Lambda<\Lambda_{\ast}$
the increasement vanishes. The evolution of the irreduciable mass
still behaves as (\ref{eq:MirrExp}). It increases exponentially at
early times and then saturates to the final value exponentially at
late times. The growth rate $\gamma_{i}$ and saturation rate $\gamma_{f}$
are shown in the lower panels of Fig.\ref{fig:LM}. $\gamma_{i}$
decreases with $-\Lambda$ and $\gamma_{f}$ increases $-\Lambda$.

{\footnotesize{}}
{\footnotesize\par}

\section{Summary and discussion}

We have studied the full nonlinear dynamics of the spontaneous scalarisation in EMS theory on spherical symmetric charged black holes in asymptotic AdS spacetimes with coupling function $f(\phi)=e^{-b\phi^{2}}$.
Solving dynamical equations governing the model numerically, we have observed detailed processes on the emergence of scalar hairs and the change of irreducible masses of black holes depending on different variables. 
Fixing the ADM mass $M=1$, we found that with sufficiently negative coupling parameter $b$, sufficiently large charge $Q$ and small enough $-\Lambda$, against a scalar perturbation an initial RN-AdS black hole will evolve into a black hole with a nontrivial scalar field outside its horizon, i.e. a hairy black hole.
The scalar hair can be characterized by the coefficient $\phi_{3}$ of the nontrivial leading order of the scalar field at infinity.
The final value $\phi_{f}$ of $\phi_{3}$ changes unsmoothly near the critical value of the coupling parameter $b_{\ast}$ and charge $Q_{\ast}$, but changes smoothly near the critcal value of cosmological constant $\Lambda_\ast$.
From the evolution of $\phi_{3}$, we learnt that there is always a zero mode when hairy black hole forms and all the higher components of the oscillating modes decay. The imaginary part of the dominant damping mode tends to zero for the black holes with larger charge or stronger coupling parameter $b$. This tells us that it takes longer period of time for a developed hairy black hole to be stabilized if the original black hole charge is bigger and the coupling is stronger.

The final irreducible mass of the developed hairy black hole changes smoothly near the critical point $b_{\ast}$,  $Q_{\ast}$ or $\Lambda_\ast$.
The irreducible mass never decreases during the evolution.
It increases exponentially at early times, approaches to and finally saturates at a finite value at late times.
For the original black hole with higher charge and stronger coupling constant $-b$, the developed hairy black hole grows earlier and takes longer time to become more massive and stabilize in the end. 
Further we have examined the phase structures in the scalarization process which is consistent with the findings in the dynamical studies that bigger original black hole charge and stronger coupling constant can make the scalarization to happen more easily.  
In the phase structure analysis, we found that the scalarization can happen more easily in the asymptotically flat spacetime than in the AdS spacetime. 
At the first sight, this result is different from what observed in the linear perturbation study \cite{Guo:2020sdu}, where it was argued that hairy black hole can be prepared more easily in AdS spacetimes. 
Considering that the endpoint of the stable configuration after scalarization can only be determined once fully non-linear numerical studies of perturbation are examined, this different observation can be understood. 
Further examinations are needed in considering non-minimal coupling between scalar field and some other source terms in the future to examine scalarizations in different spacetimes.

The coupling function we adopted in this paper is $f(\phi)=e^{-b\phi^{2}}$, which can accommodate the RN-AdS black hole solution and transform it into hairy black hole. 
The form of the coupling function is very important in the spontaneous scalarisation discussion. 
In the future, this work can be generalized to study the effect of some other coupling functions, especially to consider the dilaton function $f(\phi)=e^{-b\phi}$. 
It is known that in Einstein-Maxwell-dilaton theory, the RN black hole is not
a solution and only hairy black hole solution exists. The nonlinear
dynamics of the black hole should be studied in more details by applying the dilaton function as the coupling and examine whether a hairy black hole can be developed from an original no hair configuration \cite{Astefanesei:2019qsg,EMD}. 
One more generalization is to consider a complex scalar coupling. 
In the model considered in this paper, there is no source for the black hole charge and it is totally determined by other fields. 
For complex scalar, there will be a source of the black hole charge \cite{Hod:1996ar,Zhang:2015dwu}. 
The dynamics would be more rich and interesting \cite{Dias:2016pma,Sanchis-Gual:2015lje,Chesler:2018txn}.
Finally, we should study the evolution of black holes in asymptotic flat and dS spacetimes. The boundary conditions are different in these two spacetimes. Works in these directions are in progress.

\section*{Acknowledgments}

Peng Liu would like to thank Yun-Ha Zha for her kind encouragement during this work. This research is supported by by National Key R\&D Program of China under Grant No. 2020YFC2201400, and the Natural Science Foundation of China under Grant No. 11690021, 11947067, 12005077, 11847055, 11905083, 11805083.

\end{document}